\documentclass[12pt,notitlepage,tightenlines,eqsecnum,floats,nofootinbib,amsmath,amssymb,aps,prd, superscriptaddress]{revtex4-2}

\usepackage[utf8]{inputenc}

\usepackage{amsmath, amsfonts, amssymb}
\usepackage{mathtools}
\usepackage{mathrsfs}
\usepackage{yhmath}
\usepackage{amssymb}
\usepackage{tensor}
\usepackage{accents}
\usepackage{tipa}

\usepackage{comment}

\usepackage{fancyhdr}
\usepackage{tikz}
\usepackage{tikz-3dplot}
\usetikzlibrary{positioning}
\usetikzlibrary{shapes.geometric}
\usepackage{hyperref}
\usepackage{graphicx}

\usepackage{adjustbox}

\usepackage{setspace}

\numberwithin{equation}{section}

\usepackage{feynmf}

\usepackage{subcaption}

\usepackage[framemethod=TikZ]{mdframed}

\usepackage[T1]{fontenc}

\usepackage[english]{babel}

\usepackage{enumerate}
\usepackage{braket}
\usepackage{stackengine}
\usepackage[english]{babel}
\usepackage{amsthm}

\usepackage[framemethod=TikZ]{mdframed}
\usepackage{xcolor}

\newcommand{\tj}[6]{ \begin{pmatrix}
  #1 & #2 & #3 \\
  #4 & #5 & #6 
 \end{pmatrix}}

\newcommand{\sj}[6]{ \begin{Bmatrix}
  #1 & #2 & #3 \\
  #4 & #5 & #6 
 \end{Bmatrix}}

\newcommand{\be}{\begin{eqnarray}}
\newcommand{\ee}{\end{eqnarray}}

\newcommand{\qmarks}[1]{``#1''}
\newcommand{\diff}{\mathrm{d}}

\stackMath
\allowdisplaybreaks

\hypersetup{%
    citecolor=teal,
    urlcolor=teal,
  colorlinks=true,
  linkcolor=purple,
  linkbordercolor={0 0 1}
}

\newcommand{\myparagraph}[1]{
    
    \vspace{2mm}
    \noindent\textbf{#1}
    \hspace{-0.5em}
    }

\begin{document}

\title{An Exactly Soluble Group Field Theory}
\author{Luca Marchetti}
\email{luca.marchetti@oist.jp}
\affiliation{Department of Mathematics and Statistics, University of New Brunswick, Fredericton, NB Canada E3B 5A3}
\affiliation{Okinawa Institute of Science and Technology Graduate University,\\
Onna, Okinawa 904 0495 Japan}
\affiliation{
Kavli Institute for the Physics and Mathematics of the Universe (WPI),\\ UTIAS, The University of Tokyo, Chiba 277-8583, Japan}
\author{Hassan Mehmood}
\email{hassan.mehmood@unb.ca}
\affiliation{Department of Mathematics and Statistics, University of New Brunswick, Fredericton, NB Canada E3B 5A3}
\author{Viqar Husain}
\email{vhusain@unb.ca}
\affiliation{Department of Mathematics and Statistics, University of New Brunswick, Fredericton, NB Canada E3B 5A3}
\date{\today}

\begin{abstract}
We present a Group Field Theory (GFT) quantization of the Husain-Kuchař (HK) model formulated as a non-interacting GFT. We demonstrate that the path-integral formulation of this HK-GFT provides a complete spinfoam model and a unique Fock representation that describes the quantum three-geometries of the HK model. These results provide a link to the canonical quantization of the HK model and demonstrate how GFTs can bridge distinct quantization schemes.
\end{abstract}
\maketitle

\tableofcontents

\section{Introduction}\label{sec:intro}
Despite significant advances, it is safe to say that a fully satisfactory theory of quantum gravity is not in sight. One way to understand this state of affairs is to trace its origins to the several methodological issues inherent in formulating a quantum theory of gravity. Being radically different from systems we know how to quantize, it requires creative insight to overcome the unique obstacles it poses, often resulting in methods and approaches that may bear only the remotest analogy to known techniques. This, combined with the fact that no observations are presently available to rule out some of the possibilities, leads to questions about which methods are correct and whether there are connections between different methods.  

In view of this, one is led to consider toy models, and ask what simple systems share some of the essential structural features of general relativity in four dimensions, so that one can study the former in isolation from the latter.
In this respect models serve a threefold purpose: (i) they may provide a testing ground for different approaches to quantum gravity, especially if exactly solvable; (ii) they can be used to illustrate possible connections between different approaches; and (iii) like the Ising model, they may have significant pedagogical utility due to their relative simplicity. 

Toy models abound in quantum gravity; some notable ones are lower dimensional gravity and Chern-Simons theory \cite{Ashtekar:1989qd, Carlip:1989nz, Moncrief:1990mk, Witten:1988hc}, $\mathrm{BF}$ theories and related topological quantum field theories \cite{Horowitz:1989ng, Husain:1990sc,Blau:1989bq}, $U(1)^3$ theory \cite{Smolin:1992wj, Bakhoda:2020ril}, and various symmetry reduced models that underlie quantum cosmology \cite{BarberoG:2010oga}. All these models are more or less ``exactly soluble''. But they also have limitations. In particular, symmetry reduced models may lack the field-theoretic subtleties of general relativity and are thus of limited value in dealing with methodological issues in quantizing nonlinear field theories. The other models are   topological, i.e. lack propagating degrees of freedom, unlike general relativity. 

However, there is another model, the so-called Husain-Kuchar (HK) model \cite{Husain:1990vz}, that does not suffer from some of these limitations. Like general relativity, it is a generally covariant theory with local degrees of freedom, but it differs from the former in that it does not possess the Hamiltonian constraint. It is this model that we study here. In particular, as we shall argue, the model is well-suited to exploring the connections between three distinct but related approaches to quantum gravity, namely canonical loop quantum gravity (LQG), spinfoams and group field theory (GFT); we shall do this by developing a GFT for the model and point out its connection with the other two approaches. 

A few remarks are in order concerning why such a study is useful. Canonical LQG is to date perhaps the most successful attempt at canonically quantizing general relativity in a background-independent manner. However, as is well known, despite significant efforts, the task is still not complete, since attempts to quantize the Hamiltonian constraint of general relativity in a background-independent way are still beset with a number of problems \cite{Nicolai:2005mc, Thiemann:2007pyv}. Experience with other areas in theoretical physics has demonstrated  that covariant or path integral methods can often help sidestep the obstacles encountered in canonical quantization. In the context of LQG, spinfoams comprise the set of techniques devoted to this endeavor \cite{Baez:1999sr, Perez:2012wv, Rovelli:2014ssa}. While spinfoams have mushroomed into an independent line of inquiry into quantum gravity, the precise relationship between spinfoams and canonical LQG still remains elusive \cite{Alexandrov:2011ab}. Two salient issues are germane here. First of all, the boundary Hilbert space in spinfoam models of general relativity (for instance, the EPRL model \cite{Engle:2007qf}) is the Hilbert space of $\mathrm{SU}(2)$ spin networks. While this coincides with the \textit{kinematical} Hilbert space in canonical LQG\footnote{See \cite{Alexandrov:2011ab} for some subtle caveats in this regard.}, the question of its relationship with the \textit{physical} Hilbert space of quantum gravity remains open; to answer this question, one would need access to the physical Hilbert space in canonical LQG, but this brings one face to face with the task of quantizing the Hamiltonian constraint in canonical LQG. 

One might wish to eschew this Herculean task by observing that there exists an independent way of realizing the dynamics of general relativity at a quantum level in spinfoams by imposing the so-called simplicity constraints, and while the states from the boundary space may not strictly come from the solution space of the Hamiltonian constraint, they still are genuine states of 3-geometry, which, as just remarked, we know how to evolve. However, this still does not rule out the possibility that a fully satisfactory treatment of the Hamiltonian constraint in the canonical theory might yield a picture of dynamics quite different from what we have in spinfoam models. That there might be reason to suspect this is suggested by the fact that the second-class simplicity constraints in, say, the EPRL model are imposed at the ``quantum'' level \cite{Engle:2007qf}, which is a radical departure from how one would proceed via canonical quantization, where second-class constraints are solved classically \cite{dirac2001lectures, Henneaux:1992ig}. Thus the precise relationship between covariant spinfoams and canonical LQG is still an interesting and open subject of study. 

In view of the fact that such a study in the context of general relativity seems to inevitably run into the problem of quantizing the Hamiltonian constraint, one might wish to take a step back and ask: can one isolate our object of study from the latter problem? This is where the HK model comes in. As we pointed out above, and review in detail below, the HK model is a generally covariant theory with local degrees of freedom that does not possess the Hamiltonian constraint. Furthermore, it is a theory of connections on an $\mathrm{SU}(2)$ bundle over spacetime. Therefore, it can be \textit{exactly} quantized using canonical LQG methods \cite{Ashtekar:1995zh}; this may be regarded as a consistent quantization of all generally covariant theories of connections with local degrees of freedom modulo gauge transformations and spatial diffeomorphisms . In other words, the HK model is an exactly soluble quantum theory from the canonical perspective; its physical Hilbert space is the  kinematical Hilbert space of LQG, and the area and volume operators of LQG are physical operators. 

Thus, from the perspective of our aim, we need only construct a spinfoam for the HK model and exhibit its relationship with the known canonical quantum theory. We already know that there will be an exact correspondence between the physical Hilbert space of the HK model and the boundary Hilbert space of the associated spinfoam model, since both would comprise of $\mathrm{SU}(2)$ spin network states. Thus one part of the problem is already solved.

However, to complete the connection, a spinfoam model for HK theory is required. Such a model was suggested in \cite{Bojowald:2009im}, by restricting the spinfoam vertex of a four-dimensional $\mathrm{BF}$ theory so that it effectively  becomes  $3$-(rather than $4$-)dimensional, as required by the fact that, canonically, there is no transverse flow to change the intrinsic geometry of a 3d hypersurface in the HK model. Nonetheless, as pointed out in \cite{Bojowald:2009im}, an explicit construction of the resulting projection operator in the continuum limit, which  would lead to a definition of a physical Hilbert state to be compared with the canonical one, is still missing.

This is the reason we turn to group field theories (GFTs) \cite{Carrozza:2024gnh, Freidel:2005qe, Krajewski:2011zzu, Oriti:2006se, Oriti:2011jm}. GFTs can be seen as quantum many-body theories of \qmarks{spacetime atoms}, and, as such offer a powerful and radically new perspective on the problem of quantizing gravity. Despite continuum notions being only understood as emergent in GFTs, there has been significant progress in recent years in the extraction of continuum (especially cosmological) physics from these models \cite{Gielen:2013kla,Gielen:2013naa,Oriti:2016qtz,Gielen:2016dss,Marchetti:2020umh,Marchetti:2020qsq,Gielen:2021vdd,Calcinari:2022iss,Gielen:2020fgi,Jercher:2021bie,Gielen:2017eco,Gerhardt:2018byq,Marchetti:2021gcv,Jercher:2023nxa,Jercher:2023kfr,Oriti:2018qty,Oriti:2023yjj,Oriti:2023mgu}. Moreover, GFTs show,  in their path-integral and Fock representations,  strong connections with spinfoams \cite{Reisenberger:1996pu,Freidel:2005qe,Oriti:2011jm,Perez:2012wv} and LQG \cite{Oriti:2013aqa,oriti_inbook,Colafranceschi:2020ern}. Thus, besides being interesting in its own right, a GFT quantization of the HK model would provide a framework that can be used as a bridge between the corresponding spinfoam and canonical quantizations.

To this end we will present a quantization of the HK model as a non-interacting GFT. The resulting HK-GFT model satisfies all the physical requirements imposed by the classical HK model (including  the absence of geometry and topology changing transitions).  

Furthermore, using tools from algebraic quantum field theory applied to GFTs \cite{Kegeles:2017ems, Kegeles:2018tyo}, we construct a Fock space for such a non-interacting GFT, and show that (i) the Fock space is unique, and (ii) is equivalent to a 
symmetric sector of the standard Fock space \cite{Oriti:2013aqa,oriti_inbook,Colafranceschi:2020ern} used widely in GFTs. On the one hand, this enables one to systematically construct kinematical LQG states within the GFT Fock space,  while on the other hand, the Feynman amplitudes of the HK-GFT correspond to the spinfoam amplitudes defined in \cite{Bojowald:2009im}. Therefore, via the intermediary bridge of a non-interacting GFT, we establish a precise link between spinfoam and canonical quantization of the HK model.


The structure of the paper is as follows. In Section \ref{sec:hk}, we review the HK model, its canonical quantization via LQG methods, and establish features of a covariant quantization that the canonical theory leads us to expect. In Sections \ref{sec:intro-gft} and \ref{sec:gft-fock}, we review group field theory with a view towards the goals of the subsequent sections. Almost all the discussion in these sections is a review of known results. Our motivation for this  is to make the paper self-contained and pedagogically useful, and thus some of the discussions are somewhat detailed. New results are presented in Section \ref{sec:hk-gft}, where we argue that a non-interacting GFT serves as a satisfactory quantization of the HK model by using a constraint to reduce a general 4d GFT action with interaction to a non-interacting GFT. We study quantization of this non-interacting theory both from a path-integral and Fock-space perspective, and observe that the formalism of algebraic group field theory developed in Ref. \cite{Kegeles:2018tyo} may be used to show that the Fock space is unique in a certain sense and is equivalent to (a subspace of) the kinematical GFT Fock space studied in Ref. \cite{Oriti:2013aqa}. Finally, in Section \ref{sec:conc} we discuss some implications of describing the quantum theory of the HK model as a non-interacting GFT.

\section{The HK model}\label{sec:hk}
Since we  refer frequently to some features of the HK model to motivate its GFT model, it is useful to summarize the model. We first provide an overview of the classical HK model (Sec.\ \ref{sec:classicalhk}), and then briefly review its canonical quantization (Sec.\ \ref{sec:hk-cq}).

\subsection{Classical HK model}\label{sec:classicalhk}
Let $M$ be a 4d spacetime manifold. Define at every point of $M$ triads $e^i_{\alpha}$ ($i\in\{1,2,3\}, \alpha\in\{0,\ldots,3\}$) and connection $A^i_{\alpha}$, both of which are valued in the Lie algebra of $\mathrm{SU(2)}$. Furthermore, the triads, considered as three one-forms, are assumed to be linearly independent. The action is \cite{Husain:1990vz}
\begin{equation}
    S = \int_M \diff^4x\,\Tilde{\epsilon}^{\alpha\beta\gamma\delta}\epsilon_{ijk}e^i_\alpha e^j_\beta F^k_{\gamma\delta} \label{eq2.1}\, ,
\end{equation}
where $F^i_{\alpha\beta} = \partial_{[\alpha}A^i_{\beta]} + \tensor{\epsilon}{^i_{jk}}A^j_\alpha A^k_\beta $ is the curvature of $A$. If the $\mathfrak{su}(2)$-valued triads and connection are replaced with $\mathfrak{so}(3,1)$-valued tetrads and connection,  the Palatini action for general relativity in 4d is recovered. The use of $\mathrm{SU(2)}$ fields has significant  consequences. 

First, the  metric on $M$ defined by 
\begin{equation}
    g_{\alpha\beta} = \delta_{ij}e^i_\alpha e^j_\beta \label{eq2.2}\, ,
\end{equation}
is degenerate in a certain direction. This is seen by noting that the vector density
\begin{equation}
    \Tilde{u}^\alpha := \Tilde{\epsilon}^{\alpha\beta\gamma\delta}\epsilon_{ijk}e^i_\beta e^j_\gamma e^k_\delta\, ,
    \label{ut}
\end{equation}
is orthogonal to the triads: $\Tilde{u}^\alpha e^i_\alpha = 0$. Therefore, $g_{\alpha\beta}\Tilde{u}^\alpha = 0$. Hence the metric is degenerate in the direction determined by $\Tilde{u}^\alpha$. Because the triads are linearly independent, there are no further degeneracies \cite{Husain:1990vz}. 

The second consequence is related to the first, and arises when we inquire about the direction determined by $\Tilde{u}^\alpha$. To this end, we must convert $\Tilde{u}^\alpha$ into a vector field. This can be accomplished \cite{Husain:1990vz} if $M$ is foliated by a set of spacelike hypersurfaces, i.e. it has the topology $\mathbb{R}\times\Sigma$, where $\Sigma$ is a Riemannian 3-manifold. Then, with coordinate $t\in \mathbb{R}$ one can define the nonzero scalar density $\Tilde{e} = \Tilde{u}^\alpha\partial_\alpha t $ and the vector field $u^\alpha = \Tilde{u}^\alpha/\Tilde{e}$. That $\Tilde{e} \ne 0$ follows from the definition (\ref{ut}) provided the triads $e_\alpha^i$ are all linearly independent. Hence $u^\alpha g_{\alpha\beta}=0$, with $g$ as defined in 
  \eqref{eq2.2};  the metric is thus degenerate off the spacelike hypersurface $\Sigma$. 

The third consequence is that on the hypersurface $\Sigma$ itself, one can define a non-degenerate 3-metric. Given an embedding $X^\alpha(x^a)$ of $\Sigma$ in $M$, $x^a$ being intrinsic coordinates on $\Sigma$, the triads $e^i_\alpha$ on $M$ may be projected into $\Sigma$:
\begin{equation}
    e^i_a = e^i_\alpha \partial_a X^\alpha\, ,
\end{equation}
and define a 3-metric on $\Sigma$ by
\begin{equation}
    q_{ab} = e^i_ae^j_b\delta_{ij}\, . 
\end{equation}

Together, these arguments show that while one can define a 3-metric on an initial data slice in $\Sigma$, the theory does not possess enough ingredients to evolve this 3-metric into a unique 4-metric on $M$. This suggests that the theory does  not have a Hamiltonian constraint. This can be verified \cite{Husain:1990vz} by a direct canonical decomposition of the action \eqref{eq2.1}. The only constraints turn out to be an $\mathrm{SU(2)}$ Gauss law and a spatial diffeomorphism constraint: 
\begin{equation}
    D_a\Tilde{E}^a_i = 0\, , \qquad F^i_{ab}\Tilde{E}^b_i = 0\, .
\end{equation}
Here $\Tilde{E}^a_i = \Tilde{\epsilon}^{abc}\epsilon_{ijk}e^j_b e^k_c$, which is conjugate to the projection $A^i_a = A^i_\alpha \partial_a X^\alpha$ of the connection on $M$ into $\Sigma$, and $F^i_{ab}$ is the curvature of $A^i_a$. These are precisely the Gauss and diffeomorphism constraints for general relativity in the Ashtekar variables. 
\myparagraph{Coupling scalar fields.}
It is possible to couple scalar fields to the HK model. We describe three possibilities. The first one \cite{Husain:1993he, Rovelli:1992vv} is adding to the action \eqref{eq2.1} the term
\begin{equation}
    \int_M \diff^4x\,  \pi \Tilde{u}^\alpha\partial_\alpha\phi\, , \label{eq2.7}
\end{equation}
where $\pi$ and $\phi$ are a pair of scalar fields. Canonical analysis then shows \cite{Husain:1993he} that $\Tilde{\pi}=\Tilde{e}\pi$ is the momentum conjugate to $\phi$, and the spatial diffeomorphism constraint picks up the contribution $\Tilde{\pi}\partial_a\phi$ from the scalar fields. Again, there is no Hamiltonian constraint. Thus, like the gravitational variables, the scalar fields are non-dynamical, as can also be verified by looking at the equation of motion for $\phi$, namely $\Tilde{u}^\alpha\partial_\alpha\phi = 0$. On the other hand, if one adds the term \cite{BarberoG:1997nrd}
\begin{equation}
    -\int_M \diff^4x\,\Tilde{\epsilon}^{\alpha\beta\gamma\delta}\delta_{ij}e^i_\alpha F^j_{\beta\gamma}\partial_\delta \phi\,, \label{eq2.8}
\end{equation}
then there is a Hamiltonian constraint; this reflects the possibility of defining a non-degenerate 4-metric by  $g_{\alpha\beta} = \pm\partial_\alpha\phi\partial_\beta\phi + \delta_{ij}e^i_{\alpha}e^j_{\beta}$; this metric is non-degenerate since $\partial_\alpha\phi$ now no longer vanishes along $\Tilde{u}^\alpha$. This immediately leads to the third way of adding a scalar field to the model. One uses $\phi$ and $e^i_\alpha$ to define the tetrad \cite{BarberoG:1997nrd}
\begin{equation}
    e^I_\alpha = (\partial_\alpha\phi, e^i_\alpha)\, ,  
\end{equation}
with inverse given by
\begin{equation}
    e^\alpha_I = \frac{1}{\det{e^I_\alpha}}(\Tilde{u}^\alpha, \Tilde{u}^\alpha_i), \qquad \Tilde{u}^\alpha_i = \Tilde{\epsilon}^{\alpha\beta\gamma\delta}\epsilon_{ijk}e^j_\beta e^k_\gamma \partial_\delta\phi\, .
\end{equation}
This then allows the addition of the following term to the action \eqref{eq2.1} plus \eqref{eq2.8}:
\begin{equation}
    \int_M \diff^4x\, (\det{e^I_\alpha})e^\alpha_I e^{\beta I}\partial_\alpha\phi\partial_\beta\phi\, .
\end{equation}

The HK model without scalar fields can be understood as a gauge-fixed version of the second scalar-field model or the third one, with the gauge choice $\phi=0$. Thus, classically, both models are equivalent \cite{BarberoG:1997nrd}. We focus here on the model without scalar fields.

\subsection{Canonical Quantization}\label{sec:hk-cq}
Let us now briefly summarize the canonical quantization of the HK model using LQG methods \cite{Ashtekar:1995zh, Husain:1990vz}. The  basic result is that the physical Hilbert space consists of (group-averaged) $\mathrm{SU(2)}$ spin networks with inner product defined by the Ashtekar-Lewandowski measure. This leads to the result that unless two spin network graphs can be made to coincide through a spatial diffeomorphism, the inner product between the states they define must be zero. Put differently, since spin networks encode 3-geometries, two spin networks have a vanishing inner product unless they describe identical 3-geometries. This yields a condition that any path integral quantization of the HK model must satisfy. To see what is this condition for the spin network quantization of the HK model, let us recall that a physical state of the HK model may be represented by the ket 
\be
 | \Gamma; j_1,j_2,\cdots j_m; I_1,I_2,\cdots I_n\rangle\,,
\ee
where $\Gamma$ is a graph embedded in a 3-manifold $\Sigma$, $j_i$ are the spin labels on its edges, and $I_i$ are the intertwiners on its nodes. Now if one is given a 4-manifold with boundary that is a disjoint union $\Sigma_i\cup\Sigma_f$ of two 3-manifolds, then the transition amplitude of the HK model must satisfy 
\be 
    \langle \Gamma; j_1,j_2,\cdots j_m; I_1,I_2,\cdots I_n  | \Gamma'; j_1',j_2,'\cdots j_m'; I_1',I_2',\cdots I_n' \rangle =  \delta_{\{j_k\},\{j_k'\}}\delta_{\{I_k\},\{I_k'\}} \label{eq2.13}\, ,
\ee 
where $\Gamma'$ is the image of $\Gamma$ under a spatial diffeomorphism. This is the quantum realization of the fact that the Hamiltonian constraint of the HK model is identically zero; it indicates that there is no interaction that can create new edges and vertices, nor change the data associated with them in the process of quantum propagation of a spin network. In a pictorial spinfoam depiction, this amounts to ``dragging'' a spin network state without changing it from an initial spatial surface to a final one.  

One of the principal aims of this paper is to show that this is the case  in a group field theory quantization, and hence by implication, in a spinfoam quantization of this model. 

\section{Review of GFT}\label{sec:intro-gft}
Broadly construed, GFTs can be understood as an attempt to take seriously the idea that spacetime has a nontrivial ``quantum'' microstructure, of which our classical understanding of gravity is only a coarse-grained and incomplete description. The attempt strives to be precise enough to yield tangible, concrete models of quantum gravity, and at the same time, broad enough to encompass and possibly compare a variety of approaches to the idea of ``quantum geometry'', including, but not limited to, LQG, spinfoams, causal dynamical triangulations \cite{Jercher:2022mky}, and so on. This requires the definition of a mathematical framework that is broad and flexible enough to capture the main physical insights of the above approaches. This is provided by the tensorial group field theory (TGFT) formalism. TGFTs are defined in terms of a complex- or real-valued tensor field $T^{AB\ldots}$ on $r$ copies of a group $G$, with dynamics characterized by combinatorially non-local interactions:
\begin{equation}\label{eqn:tgft}
    S[T,T^*] =\mathrm{tr}\left(T^*, KT\right)+\sum_\gamma\lambda_\gamma U_\gamma[T^*,T]+\text{c.c}\,. 
\end{equation}
In the above equation, the first term is a \qmarks{kinetic} term, where we have defined the inner product $(\cdot,\cdot)$ as
\begin{equation}
    \left(T,T'\right):=\int_{G^r} \diff g_I\, T(g_I)T'(g_I)\,,
\end{equation}
with $(g_1,\ldots,g_r):=(g_I)\in G^r$,  and $\diff g_I=\prod_{i=1}^r\diff g_i$, where $\diff g_i$ is an appropriate measure on $G$. The second term in equation \eqref{eqn:tgft} is an \qmarks{interaction} term. It is characterized by combinatorially non-local contractions of the tensor field data (tensor indices $AB\dots$, and group variables $g_I$) dictated by patterns represented by the graphs $\gamma$, each vertex of which represents a power of $T$ (or $T^*$). GFTs are then TGFTs that admit a quantum gravitational interpretation, i.e., whose group manifold can be associated with quantum geometric data, and  whose interactions are (often, but not necessarily) characterized by combinatorial patterns associated with the gluing of $(d-1)$-dimensional cells to form a $d$-dimensional one.

\subsection{Representations of the group field}\label{sec:gft-reps}

To make the above idea concrete, we provide a concrete illustration by focusing on a complex-valued scalar field on copies of $\mathrm{SU}(2)$, which is intended to represent a (quantum) tetrahedron. (We will see below in  Section\ \ref{sec:gft-amps} how to specify interactions that describe the gluing of $5$ such tetrahedra to form the simplest building block of a $4$-dimensional spacetime --- a $4$-simplex.) 

There are three useful representations of the group field, the group representation, the spin representation and the Lie algebra representation. For the purposes of this paper, we shall require only the first two, and hence it is useful to review them.

\myparagraph{Group representation.} The group field is defined, in general, by the complex-valued function 
\be
\phi : \mathrm{SU}(2)^4\to\mathbb{C}\, , 
\ee
with the condition that it be invariant under the right action of $\mathrm{SU}(2)$:
\begin{equation}
    \phi(g_1,g_2,g_3,g_4)=\phi(g_1h,g_2h,g_3h,g_4h)\,, \quad \forall h, g_i \in \mathrm{SU}(2)\, .\label{right-invar}
\end{equation}
The reason $\phi$ is a function of four group elements is that it is designed to represent the ``wavefunction'' associated to a single tetrahedron, with each of its faces associated with a group element $g$. Any field $\chi(g_1,\cdots,g_4)$ can be converted into a right-invariant one by means of a projection operator,
\begin{equation}
   \phi(g_1,\ldots,g_4):= \mathcal{P}\chi(g_1,\dots, g_4) = \int \diff h\,\chi(g_1h,g_2h,g_3,g_4h) \label{proj-op}\, ,
\end{equation}
where the integration above is done using normalized Haar measure on $\mathrm{SU}(2)^4$.
 
\myparagraph{Spin representation.} 
  By the Peter-Weyl theorem, one can expand the fields $\phi: \mathrm{SU}(2)^4\to\mathbb{C}$ in terms of the matrix elements $D^{j}_{mn}(g)$ of spin-$j$ representations of $\mathrm{SU}(2)$:
\begin{align}
    \phi(g_1, g_2, g_3, g_4) &= \sum_{j_i, m_i, n_i}\phi^{j_1\ldots j_4}_{m_1\ldots m_4;n_1\ldots n_4}D^{j_1}_{m_1n_1}(g_1)\cdots D^{j_4}_{m_4n_4}(g_4) \nonumber\\
                &= \sum_{J,M,N}\phi^{J}_{MN}D^{J}_{MN}(g_I)\, ,\label{spinrep1}
\end{align}
where we have introduced the notation $J=(j_1,j_2,j_3,j_4)$, $D^J_{MN}(g_I) = D^{j_1}_{m_1n_1}(g_1)\cdots D^{j_4}_{m_4n_4}(g_4)$, etc. By virtue of \eqref{orth-3} and \eqref{proj-op}, the above equation can be written as
\begin{equation}
    \phi(g_I) = \sum_{JMNj}\varphi^{Jj}_{M}I^{Jj}_{N}D^{J}_{MN}(g_I)\sqrt{d_J}, \quad \varphi^{Jj}_{M} \equiv \varphi^{j_1j_2j_3jm_4;j}_{m_1m_2m_3m_4} \equiv \sum_{N}\phi^{J}_{MN}I^{Jj}_{N}\,, \label{spinrep2}
\end{equation}
where $d_J := 2J+1 := (2j_1+1)\cdots(2j_4+1)$, and we have introduced the so-called intertwiners $I^{Jj}_{N}$, which span the invariant part of $V_{j_1}\otimes\cdots\otimes V_{j_4}$; $V_{j_i}$ is the vector space of the spin-$j_i$ representation. Explicitly, they are given by 
\begin{equation}
    I^{Jj}_{N} = \sqrt{d_j}\sum_{n}(-1)^{j-n}\tj{j_1}{j_2}{j}{n_1}{n_2}{n}\tj{j}{j_3}{j_4}{-n}{n_3}{n_4}\, \label{inter}
\end{equation}
where the objects with round brackets are the Wigner 3$j$-symbols used in the theory of quantum angular momentum. They are zero unless the magnetic quantum numbers in them sum to zero (e.g. $n_1+n_2+n=0$ above) and the $j$ quantum numbers satisfy the triangle inequality (e.g. $|j_1-j_2|\leq j \leq j_1+j_2$). Furthermore, they have a number of properties and calculations involving them can be done using an elaborate diagrammatology. For details, see Refs. \cite{1962mata.book.....Y, 1988qtam.book.....V, Martin-Dussaud:2019ypf}. We only note a few important identities among $D^{j}_{mn}(g)$ and $I^{Jj}_{M}$ that we will require.

First, note some orthogonality relations satisfied by the Wigner matrices \cite{Ooguri:1992eb}, 
\begin{align}
    \int dh \prod_{i=1}^{2}D^{j_i}_{m_in_i}(h) &= \delta^{j_1j_2}\frac{(-1)^{j_1-m_1}(-1)^{j_2-n_1}}{\sqrt{(2j_1+1)(2j_2+1)}}\delta_{m_1,-m_2}\delta_{n_1,-n_2} \label{orth-1}\,,\\
    \int dh \prod_{i=1}^{3}D^{j_i}_{m_in_i}(h)& = \tj{j_1}{j_2}{j_3}{m_1}{m_2}{m_3}\tj{j_1}{j_2}{j_3}{n_1}{n_2}{n_3} \label{orth-2}\,,\\
    \int dh \prod_{i=1}^{4}D^{j_i}_{m_in_i}(h) &= \sum_{j}I^{j_1j_2j_3j_4j}_{m_1m_2m_3m_4}I^{j_1j_2j_3j_4j}_{n_1n_2n_3n_4}\,, \label{orth-3}
\end{align}
Also important is the Peter-Weyl representation of the Dirac delta function on $\mathrm{SU(2)}$ (defined to be nonzero whenever its argument is equal to the unity element in $\mathrm{SU(2)}$): 
\begin{equation}
    \delta(g^{-1}h) = \sum_{jmn}d_j\Bar{D}^{j}_{mn}(g)D^{j}_{mn}(h) = \sum_{jmn}d_j(-1)^{m-n}D^{j}_{-m-n}(g)D^{j}_{mn}(h) \label{peter-weyl-delta}\, 
\end{equation}
where the second equality comes from the fact that
\begin{equation}
    \overline{D}^{j}_{mn}(g) = (-1)^{m-n}D^{j}_{-m-n}(g)\label{reality-wigner}\,.
\end{equation}

We learn a number of things from \eqref{spinrep2}. First, if the field $\phi$ is assumed to be real, then since $D^{J}_{MN}(g_I) = (-1)^{\sum_{i=1}^{4}m_i-n_i}D^{J}_{-M-N}$, $I^{Jj}_{N}=(-1)^{\sum_{i=1}^4 j_i}I^{Jj}_{-N}$ and $\sum_{i=1}^{4}n_i=1$ for an intertwiner,
\begin{equation}
    \bar{\varphi}^{Jj}_M = (-1)^{\sum_{i=1}^{4}j_i-m_i}\varphi^{Jj}_{-M}, \label{reality-coeff}
\end{equation}
Second, we can see that a GFT field $\phi(g_1,\ldots,g_4)$ is composed of linear combinations of $D^J_{MN}(g_I)I^{Jj}_N\sqrt{d_J}$, with coefficients $\varphi^{Jj}_M$, much like a mode expansion in usual field theory; the modes now are $\varphi^{Jj}_M$, labeled by spin, angular momentum and intertwiner data. We can also define an $\mathcal{H}_v$ as the invariant subspace of the tensor product of the spin-$j$ Hilbert spaces $\mathcal{H}^{(j)}$ associated with the spin decomposition of a function $f\in L^2(\mathrm{SU}(2))$, i.e., $\mathcal{H}_v=\text{Inv}(\otimes_{i=1}^4\mathcal{H}^{(j_i)})$. The field modes $\varphi^{Jj}_{N}$ can also be naturally associated with a tetrahedron: the tetrahedron itself carries an intertwiner label $j$ and its four faces are labeled by $(j_1,m_1),\ldots, (j_4,m_4)$ respectively; the closure of the faces now translates to the presence of an intertwiner in the definition of $\varphi^{Jj}_M$, which ensures $m_1+\cdots +m_4=0$. We call such a tetrahedron an \textit{open tetrahedron}; the  GFT field is a linear combination of such open tetrahedra.

It is also worth pointing out that every tetrahedron has a dual graph: replace the tetrahedron by a vertex at its center and four edges connecting the vertex and the four faces of the tetrahedron respectively. Accordingly, fields and field modes can equivalently be associated with these dual graphs. In particular, in the case of the field modes, we will call the dual graph a dual \textit{open spin network}. 

To summarize, a GFT field  can be graphically represented by a tetrahedron, or equivalently by its dual open spin network---a node with $4$ outgoing half-links. The tetrahedron, or its dual open spin network, is decorated with group or spin data, depending on the representation chosen.

\subsection{Group field theory amplitudes and simplicial path-integrals}\label{sec:gft-amps}
We have seen that a GFT field can naturally be associated with a tetrahedron and its dual spin network, and the association works at the level of group and representation theory. The next step is to construct a GFT action. The choice of the action is motivated by our desire to describe by means of it a quantization of a classical theory of geometries.

As an illustration and in anticipation of future use, let us focus on the case of 4d $\mathrm{BF}$ theory, which is a topological field theory formulated on a 4-manifold (see Appendix \ref{sec:appC} for a review). Its GFT is described by the Ooguri action \cite{Ooguri:1992eb}:
\begin{align}
    S_O &= \frac{1}{2}\int \diff g_1\cdots \diff g_4\ \phi^2(g_1, \ldots, g_4)\nonumber\\  &\quad+\frac{\lambda}{5!}\int \diff g_1\cdots \diff g_{10} \ \phi(g_1, g_2, g_3, g_4)\phi(g_4, g_5, g_6, g_7)\phi(g_7, g_3, g_8, g_9)\nonumber \\
    &\qquad\qquad\qquad \times\   \phi(g_9, g_6, g_2, g_{10})\phi(g_{10}, g_8, g_5, g_1)\, . \label{Oaction}
\end{align}
Here the field $\phi$ is taken to be real-valued. The geometric rationale for this action is as follows. As explained above, $\phi(g_1,g_2,g_3,g_4)$ represents a tetrahedron with its four faces labeled by the $g_i$; the kinetic term describes two tetrahedra with all faces glued together in pairs. The interaction term  describes five tetrahedra, each of which shares a face with every other one; the resulting structure is a 4-simplex (like 4 triangles joined together via pairwise edge-to-edge gluings to form a 3-simplex) -- the order of the arguments of fields in the interaction term dictates which face is glued to which, so as to result in a 4-simplex (again with comparison to  the edges of the triangles glued in a specific way to produce a 3-simplex). 

\begin{figure}[ht]
    \centering
    \begin{subfigure}{.4\linewidth}
    \begin{adjustbox}{trim=0 0 0 3cm}
        \begin{tikzpicture}
    \def\L{2.5}
    \def\l{.5/2*3}
    \node (x1) at (\L,-\l) {$x_1$};
    \node (x2) at (\L,-\l/3) {$x_2$};
    \node (x3) at (\L,\l/3) {$x_3$};
    \node (x4) at (\L,\l) {$x_4$};
    \begin{scope}[rotate=180]
        \node (y4) at (\L,-\l) {$y_4$};
        \node (y3) at (\L,-\l/3) {$y_3$};
        \node (y2) at (\L,\l/3) {$y_2$};
        \node (y1) at (\L,\l) {$y_1$};
    \end{scope}
    \draw (x1) -- (y1);
    \draw (x2) -- (y2);
    \draw (x3) -- (y3);
    \draw (x4) -- (y4);
    \draw[fill=white] (-\l,-1.5*\l) rectangle (\l,1.5*\l);
    \node at (0,0) {$\tau$};
    \end{tikzpicture}
    \end{adjustbox}
    \caption{Propagator}
    \end{subfigure}
    \begin{subfigure}{.4\linewidth}
        \begin{tikzpicture}
    \def\L{2.5}
    \def\l{.5/2*3}
    \node (y9) at (\L,-\l) {$y_9$};
    \node (y6) at (\L,-\l/3) {$y_6$};
    \node (y2) at (\L,\l/3) {$y_2$};
    \node (x10) at (\L,\l) {$x_{10}$};
    \begin{scope}[rotate=72]
        \node (y10) at (\L,-\l) {$y_{10}$};
        \node (y8) at (\L,-\l/3) {$y_8$};
        \node (y5) at (\L,\l/3) {$y_5$};
        \node (y1) at (\L,\l) {$y_1$};
    \end{scope}
    \begin{scope}[rotate=2*72]
        \node (x1) at (\L,-\l) {$x_1$};
        \node (x2) at (\L,-\l/3) {$x_2$};
        \node (x3) at (\L,\l/3) {$x_3$};
        \node (x4) at (\L,\l) {$x_4$};
    \end{scope}
    \begin{scope}[rotate=3*72]
        \node (y4) at (\L,-\l) {$y_4$};
        \node (x5) at (\L,-\l/3) {$x_5$};
        \node (x6) at (\L,\l/3) {$x_6$};
        \node (x7) at (\L,\l) {$x_7$};
    \end{scope}
    \begin{scope}[rotate=4*72]
        \node (y7) at (\L,-\l) {$y_7$};
        \node (y3) at (\L,-\l/3) {$y_3$};
        \node (x8) at (\L,\l/3) {$x_8$};
        \node (x9) at (\L,\l) {$x_9$};
    \end{scope}
    \draw (x1) to [bend right] (y1);
    \draw (x2) to [bend right = 15] (y2);
    \draw (x3) to [bend left = 15] (y3);
    \draw (x4) to [bend left] (y4);
    \draw (x5) to [bend right = 15] (y5);
    \draw (x6) to [bend left = 15] (y6);
    \draw (x7) to [bend left] (y7);
    \draw (x8) to [bend left = 15] (y8);
    \draw (x9) to [bend left] (y9);
    \draw (x10) to [bend left] (y10);
    \end{tikzpicture}
    \caption{Vertex}
    \end{subfigure}
    \caption{Vertex and propagtor for the Ooguri model.}
    \label{fig:vertex-prop}
\end{figure}
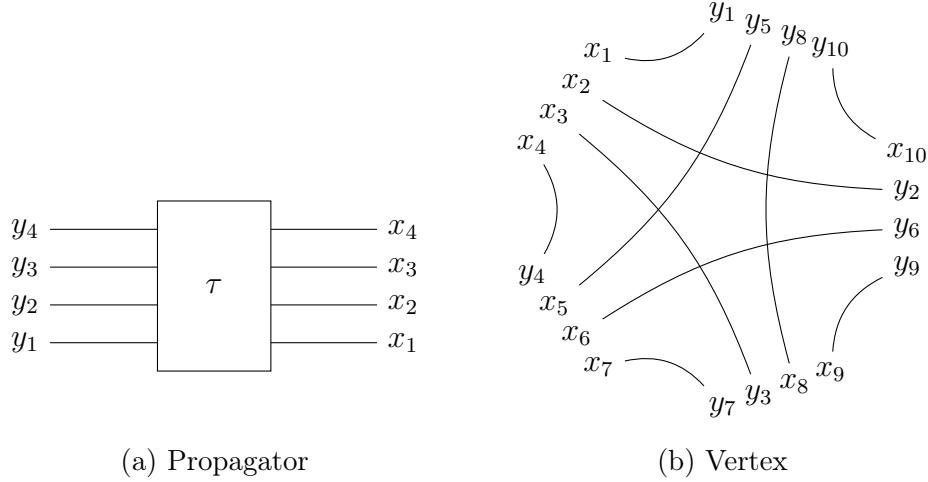

This shows that the vertex of this theory is a 4-simplex, and a propagator is either the gluing together of two tetrahedra coming from two distinct 4-simplices or the same 4-simplex (self-intersections); see Fig \ref{fig:vertex-prop}. Thus the the Feynman diagrams of this theory are collections of 4-simplices (vertices) glued together along their boundary tetrahedra through the propagator
\begin{equation}
    P(g_I, h_I) =\int \diff s \,\prod_{i=1}^{4}\delta(g_ish^{-1}_i), \label{free-prop}
\end{equation}
which has been written in a gauge-invariant manner. The propagator describes gluing of two tetrahedra labeled by $(g_I)$ and $(h_I)$ respectively. The objects resulting from gluing tetrahedra on the boundary of 4-simplices are arbitrary cellular complexes, i.e. objects composed of 0, 1, 2, $\ldots$, $n$ cells or simplices, together with a prescription for gluing some $i$-cells in the collection across $(i-1)$-cells on their boundaries. This means that the partition function 
\begin{equation}
    Z = \int D\phi \exp{(-S_O[\phi])}\, , \label{PI}
\end{equation}
expanded in powers of $\lambda$ would scale with the number of vertices in a Feynman diagram to give 
\begin{equation}
    Z = \sum_C \frac{\lambda^{N_4(C)}}{N_{\text{sym}}(C)}Z_C \label{eq3.17}\, ,
\end{equation}
where the sum is over all possible cellular complexes $C$ dual to the Feynman graphs of the theory, $Z_C$ is the quantum amplitude associated with the complex $C$ (where  complexes with the same number $N_4(C)$ of $4$-simplices are weighted by the same power of $\lambda$ from the perturbative expansion of the action,), and $N_{\text{sym}}(C)$ is the order of the symmetry group of $C$. A cellular complex in which no two glued tetrahedra come from the same 4-simplex represents a triangulation of a 4-manifold. Thus the sum in \eqref{PI} contains in particular a sum over all combinatorial 4-manifolds. $Z$ would represent a generating functional for $\mathrm{BF}$ simplicial path-integrals if the amplitudes $Z_C$ can be identified with a $\mathrm{BF}$ path-integral discretized over the complex $C$. This is what we show below, both in the group and in the spin representation. 

 \myparagraph{Spin representation of the amplitudes.}
From \eqref{spinrep2}, and the assumption of permutation invariance of $\phi(g_1,g_2,g_3,g_4)$ in any three of its arguments \cite{Ooguri:1992eb}, it can be shown that the action \eqref{Oaction} becomes
\begin{align}
    S_O &= \frac{1}{2}\sum_{JMj}|\varphi^{Jj}_{M}|^2 - \frac{\lambda}{5!}\sum_{\{j_i,l_i,m_i,n_i\}}(-1)^{\sum_{i=1}^{10}(j_i+m_i)}(-1)^{\sum_{i=1}^{5}(l_i+n_i)}\begin{Bmatrix}
 l_1& l_2 & l_3 &l_4  &l_5  \\
j_1 & j_2 & j_3 & j_4 & j_5 \\
 l_{10}& l_9 & l_8 & l_7 & l_6 \\
\end{Bmatrix} \nonumber \\
&\qquad\qquad\qquad\times \varphi^{j_2l_2l_4j_3;l_3}_{m_2n_2-n_4m_3}\varphi^{j_4l_4l_6j_5;l_5}_{m_4n_4-n_6m_5}\varphi^{j_2l_6l_8j_1;l_7}_{-m_2n_6-n_8m_1}\varphi^{j_3l_8l_{10}j_5;l_9}_{-m_3n_8-n_10-m_5}\varphi^{j_4l_{10}l_2j_1;l_1}_{-m_4n_{10}-n_2-m_1} \nonumber \\
&= \frac{1}{2}\sum_{JMj}|\varphi^{Jj}_{M}|^2 + S_{\text{int}},
\end{align}
where the object in the big curly braces is the $15j$ symbol of the third kind, which is a specific sum over products of $3jm$ symbols, also encountered in angular momentum theory \cite{ 1988qtam.book.....V, 1962mata.book.....Y}. This enables one to write the amplitude $Z_C$ associated with a particular complex $C$ explicitly. To this end, we should first of all determine all the nonvanishing components of the propagator, i.e. the two-point function of the theory. They are encoded in  \cite{Ooguri:1992eb}
\begin{align}
    \braket{\varphi^{j_1j_2j_3j_4j}_{m_1m_2m_3m_4}\overline{\varphi}^{j_1j_2j_3j_4j}_{m'_1m'_2m'_3m'_4}} &= \frac{1}{9}\prod_{i=1}^{4}\delta_{m_i,m'_i}\,,\\
    \braket{\varphi^{j_1j_2j_3j_4j}_{m_1m_2m_3m_4}\overline{\varphi}^{j_3j_1j_2j_4j'}_{m'_3m'_1m'_2m'_4}} &= \frac{1}{9}\prod_{i=1}^{4}\delta_{m_i,m'_i}\sqrt{(2j+1)(2j'+1)}\sj{j_1}{j_2}{j'}{j_3}{j_4}{j}\,,
\end{align}
and similar relations obtained by cyclic permutations of $(j_1,j_2,j_3,j_4)$; here $\braket{(\cdots)}$ represents the expectation value of $(\cdots)$ with respect to the  Gaussian measure 
\be
\prod_{JMj}\diff \varphi^{Jj}_M\exp{\left(-\frac{1}{2}\sum_{JMj}|\varphi^{Jj}_M|^2\right)}\, .
\ee
and the object in the curly braces is the $6j$ symbol from the theory of angular momentum \cite{1962mata.book.....Y, 1988qtam.book.....V, Martin-Dussaud:2019ypf}. This means that whenever a propagator connects two vertices or connects a vertex to itself, one either picks a delta function or a delta function times a $6j$-symbol. Thus, since we know that the propagator represents propagation of tetrahedra, each tetrahedron in a Feynman graph generally carries a $6j$-symbol. On the other hand, $S_{\text{int}}$ is a 4-simplex with which one can associate a $15j$-symbol \cite{Baez:1999sr}. Therefore, schematically, one should have \cite{Ooguri:1992eb}  
\begin{equation}
    Z_C = \sum_j \prod_{f: \text{2-simplices}}(2j_f + 1) \prod_{\text{3-simplices}}\{6j\} \prod_{\text{4-simplices}}\{15j\} \label{eq3.18}\, .
\end{equation}
This is similar in form to the amplitude from a spinfoam quantization of $\mathrm{BF}$ theory on $C$, if $C$ is a combinatorial manifold (Appendix \ref{sec:appC}). In other words, the partition function \eqref{PI} of a GFT contains a sum over combinatorial manifolds of the spinfoam amplitudes associated with those manifolds. In this way, a GFT helps bypass the delicate question of triangulation independence in spinfoams, which can be answered in the affirmative only for topological theories like $\mathrm{BF}$ theory \cite{Baez:1999sr} (where the sum over combinatorial manifolds reduces to a sum over distinct topologies). One can thus see a GFT as a completion of its underlying triangulation-dependent spinfoam model. 

\myparagraph{Amplitudes in the group representation.} Once we have obtained $Z_C$ in the spin representation, it is a matter of using appropriate identities from the harmonic theory of $\mathrm{SU(2)}$ \cite{Martin-Dussaud:2019ypf} to pass over to the group representation. To this end, note first that every face $f$ in $C$ is shared by a finite number of tetrahedra, say $t_{1}, \ldots, t_{n}$. Since each tetrahedron has its faces labeled by group elements, a face $f$ has associated with it different group elements $h_{f(t_1)}, \ldots, h_{f(t_n)}$ coming from the tetrahedra to which it belongs. Keeping this in mind and using \eqref{orth-1}-\eqref{orth-3} and the definition of the $6j$- and $15j$-symbols in terms of the $3jm$-symbols, $Z_C$ can be written as \cite{Ooguri:1992eb}
\begin{equation}
    Z_C = \sum_{j}\int \prod_{t:\text{3-simplices}} \diff h_t\prod_{f:\text{2-simplices}} (2j_f+1)\, \text{tr}(D^{j_{f(t_1)}}(h_{f(t_1)})\cdots D^{j_{f(t_n)}}(h_{f(t_n)}))\, ,
\end{equation}
But this reduces using \eqref{peter-weyl-delta} to
\begin{equation}
    Z_C = \int \prod_{t:\text{3-simplices}}\diff h_t\prod_{f:\text{2-simplices}}\, \delta(h_{f_{(t_1)}}\cdots h_{f_{(t_n)}})\, . \label{eq3.20}
\end{equation}
Recalling that in $\mathrm{BF}$ theory, tetrahedra sharing a face in $C$ correspond to the edges surrounding a face in the dual 2-complex of $C$ (see Appendix \ref{sec:appC}), this reproduces again the spinfoam amplitude for a $\mathrm{BF}$ theory on $C$, this time in the group representation; the $h_{f_{(t_i)}}$ are the holonomies along the edge dual to the tetrahedron $t_i$. 

\subsection{Aside on real and complex GFT fields}
In reviewing the Ooguri model, we have used a real-valued GFT field. But this is not necessary. One can also use a complex-valued field $\phi:SU(2)^4\to\mathbb{C}$ to define an action (provided its is real); the kinetic term would be of the form $\phi\bar{\phi}$. On the other hand, there can be more than one interaction term of the form appearing in the Ooguri action \eqref{Oaction}, depending on how one wants to convolve the field and its conjugate in the vertex of the theory. For instance, one possibility is
\begin{equation*}
    \int \diff g_1\cdots \diff g_{10} (\phi_{1234}\phi_{4567}\phi_{7389}\phi_{96210}\phi_{10851} + \bar{\phi}_{1234}\bar{\phi}_{4567}\bar{\phi}_{7389}\bar{\phi}_{96210}\bar{\phi}_{10851})\, ,
\end{equation*}
where we have used the shorthand $\phi(g_1,g_2,g_3,g_4)\equiv\phi_{1234}$, etc. Similarly, another possibility could be permutations of terms of the form $\bar{\phi}\phi^4 + \phi\bar{\phi}^4$, and so on. 

There are now two kinds of tetrahedra in the theory, those labeled by $\phi$ and those labeled by $\bar{\phi}$. The kinetic term $\phi\bar{\phi}$ now describes overlapping of distinct types of tetrahedra. In particular, the propagator of the theory is nonzero only for propagation between $\bar{\phi}$ and $\phi$ tetrahedra, i.e. $\braket{\phi\phi}=\braket{\bar{\phi}\bar{\phi}}=0$. Furthermore, since the interaction terms still represent 4-simplices, the partition function of the theory can still be organized as a sum over combinatorial structures, including 4-manifolds. The only difference is that 4-simplices are glued along $\phi$ and $\bar{\phi}$, rather than along $\phi$ and $\phi$ or $\bar{\phi}$ and $\bar{\phi}$, tetrahedra. 

These considerations show that a complex-valued GFT field gives a geometrically sensible theory, as does the real-valued field. However, the former is evidently more general. It also provides an arena wherein a Fock space for a GFT can be postulated. This in turn facilitates comparison with canonical LQG. We now turn to these matters.

\section{GFT Fock space}\label{sec:gft-fock}

\subsection{Fock space for a complex GFT field}

Fock quantization of GFT has been developed in a series of papers in Refs.  \cite{Oriti:2013aqa, Colafranceschi:2020ern, Kegeles:2017ems} (and following different approaches in \cite{Mikovic:2001yg} and \cite{Perez:2001gja}); for a review, see \cite{Gielen:2024sxs}. To begin with, we take a complex-valued GFT field\footnote{We specialize to four copies of $\mathrm{SU}(2)$ for simplicity. The results generalize to any number of copies \cite{Oriti:2013aqa}.} $\phi: SU(2)^4\to \mathbb{C}$ and promote it and its conjugate to operators $\hat{\phi}$ and $\hat{\phi}^\dagger$ by imposing the following commutation rules.
\begin{equation}
    [\hat{\phi}(g_I),\hat{\phi}(h_I)]=[\hat{\phi}^\dagger(g_I),\hat{\phi}^\dagger(h_I)]=0, \quad [\hat{\phi}(g_I), \hat{\phi}^\dagger(h_I)] = \int \diff s\,\prod_{i=1}^{4}\delta(g_ish^{-1}_i). \label{canon-comm-complex}
\end{equation}
Here we have used a gauge-invariant delta function to ensure consistency with the right-invariance of the fields. This is the algebra of creation and annihilation operators. The motivation for this comes from thinking of a GFT as a many-body theory of quanta of space \cite{Oriti:2013aqa}. That is, $\hat{\phi}^\dagger(g_I)$ and $\hat{\phi}(g_I)$ are creation and annihilation operators, respectively, creating and destroying a tetrahedron labeled by $(g_I)$. There is thus a vacuum $\ket{0}$ which is annihilated by $\hat{\phi}$ and on which repeated application of $\hat{\phi}^\dagger$ gives rise to a Fock space. 

The geometric interpretation becomes more transparent in the spin representation. Defining $C^{Jj}_{M}(g_I)\equiv \sum_{N}\sqrt{d_J}I^{Jj}_{N}D^{J}_{MN}(g_I)$, we write (cf. \eqref{spinrep1})
\begin{equation}
    \Hat{\phi}(g_I) = \sum_{JjM} \Hat{a}^{Jj}_M\, C^{Jj}_{M}(g_I), \quad \hat{\phi}^\dagger(g_I) = \sum_{JjM}\Hat{a}^{\dagger Jj}_{M}\, \bar{C}^{Jj}_M(g_I)\,  \label{eq3.23}
\end{equation}
where the operators $\hat{a}^{Jj}_M$ and $\hat{a}^{\dagger Jj}_M$ satisfy
\begin{equation}
    [\Hat{a}^{Jj}_M, \Hat{a}^{Kk}_N] = [\Hat{a}^{\dagger Jj}_M, \Hat{a}^{\dagger Kk}_N] = 0,\quad [\Hat{a}^{Jj}_M, \Hat{a}^{\dagger  Kk}_N] = \delta^{J,K}\delta_{M,N}\delta_{j,k}\, . \label{eq3.43}
\end{equation}
Recall that the tetrahedra associated to field configurations have four-valent dual open spin networks, and in particular, the dual spin networks labeled by field modes carry spins, magnetic quantum numbers and an intertwiner label.  With this interpretation, a creation operator $a^{\dagger Jj}_{M}$ acts on the vacuum state $\ket{0}$ to create a 4-valent vertex whose four (in principle indistinguishable) links are labeled by $(j_1,m_1), (j_2,m_2), (j_3,m_3), (j_4, m_4)$ and the vertex itself is labeled by an intertwiner quantum number $j$; we represent these states by $\ket{J,M,j}$. The annihilation operator $A^{Jj}_M$ destroys such vertices. Thus, the one-particle states are these 4-valent open spin networks belonging to $\mathcal{H}_v$ \cite{Oriti:2013aqa}. A GFT Fock space $\mathcal{F}_{\text{GFT}}$ can then be constructed in the usual manner by repeated action of the creation operator over the Fock vacuum. In the occupation number basis, the states generated in this way 
are labeled by the numbers of distinct 4-valent open spin networks corresponding to distinct labels $(J,M,j)$ and provide a convenient basis for $\mathcal{F}_{\text{GFT}}$. For instance, the state $\ket{n_{JMj}, n_{KNk}}$ contains $n_{JMj}$ and $n_{KNk}$ 4-valent open spin networks labeled by $(J,M,j)$ and $(K,N,k)$ respectively, and no other vertices. The action of, for instance, $a^{\dagger Jj}_{M}$ and $a^{Jj}_{M}$ on these states will be
\begin{align}
    a^{\dagger Jj}_{M}\ket{n_{JMj}, n_{KNk}} &= \sqrt{n_{JMj}+1}\ket{n_{JMj}+1, n_{KNk}},\\ a^{ Jj}_{M}\ket{n_{JMj}, n_{KNk}} &= \sqrt{n_{JMj}}\ket{n_{JMj}-1, n_{KNk}}\, .
\end{align}
The inner product between one-particle states is the natural one induced from the standard $L^2$ inner product with respect to Haar measure between one-particle wave functions $C^{Jj}_{M}(g_I)$. That is, from \eqref{orth-1}-\eqref{orth-3}, we have
\begin{equation}
    \int \diff g_I \, C^{Jj}_{M}(g_I) \bar{C}^{Kk}_{N}(g_I) = \delta^{J,K}\delta_{M,N}\delta_{j,k}.
\end{equation}
Thus, we define
\begin{equation}
    \braket{J,M,j|K,N,k} = \braket{0|a^{Jj}_{M}a^{\dagger Kk}_{N}|0} \equiv \delta^{J,K}\delta_{M,N}\delta_{j,k}
\end{equation}

Explicitly, we can write down the Fock space as
\begin{equation}
    \mathcal{F}_{\text{GFT}}=\bigoplus_{V=0}^\infty\mathcal{H}_{V},\quad \quad \mathcal{H}_{V} \equiv \mathrm{sym}(\mathcal{H}_v\underbrace{\otimes\cdots\otimes}_{V \text{ times}}\mathcal{H}_v)~, \label{gft-fock-sp}
\end{equation}
where $\mathcal{H}_v = L^2(\mathrm{SU(2)^4}/\mathrm{SU(2)})$ is the space of right-invariant functions on $\mathrm{SU(2)}^{4}$. For later purposes, it is also convenient to define a \textit{pre-Fock} space, which is simply the unsymmetrized direct product over $\mathcal{H}_v$, i.e.
\begin{equation}
    \tilde{\mathcal{F}}_{\text{GFT}}=\bigoplus_{V=0}^\infty\left(\mathcal{H}_v\underbrace{\otimes\dots\otimes}_{V \text{ times}} \mathcal{H}_v\right)\, . \label{gft-pre-fock}
\end{equation}
It is worth emphasizing that the Fock space presented above is not a construction from first principles. Rather, it has been simply postulated. The analogy is with the Fock spaces encountered in non-relativistic quantum field theory of many-body systems \cite{Oriti:2013aqa}. How far is such an analogy justified within a quantum gravity context is open for debate. For now, we will simply work with this Fock space, without asking where it came from or how can it be derived solely within a GFT framework. We will return to this question later in Section \ref{sec:hk-gft} when we study the GFT of the HK model. 

\subsection{The kinematical LQG Hilbert space and the GFT Fock space}
As we have seen above, open 4-valent spin network vertices can be thought of as the dual graphs of a tetrahedron, and by acting on the vacuum with the creation operators and contracting the magnetic quantum numbers on different operators, one can construct arbitrary spin networks. This suggests that it is possible to construct in the GFT Fock space arbitrary spin networks in the diffeo-invariant Hilbert space of canonical LQG $\mathcal{H}_{\text{LQG}}$. This is spelled out in more detail in \cite{Oriti:2013aqa, Colafranceschi:2020ern, Sahlmann:2023plc}. 

\myparagraph{Kinematical states.}
Take an arbitrary kinematical spin network state, described in the following way. Consider a graph $\Gamma$ with $V$ vertices, labeled with lowercase Latin letters such as $i=1,\ldots,V$. For simplicity, let each vertex be $d$-valent, the links surrounding it denoted by Greek letters like $\alpha =1,\cdots,d$. Every edge in the graph connects two vertices together. To make the connectivity explicit, we denote by $(i\alpha,j\beta)$ the edge that goes from the $\alpha$th link at the $i$th vertex to the $\beta$th link at the $j$th vertex. Each such edge carries a spin label $j^{\alpha\beta}_{ij}$, and to ensure guage invariance, each vertex $i$ carries an intertwiner $I_i$ between the Hilbert spaces corresponding to the spins on the links surrounding $i$. In this way, we get the spin network state
\begin{equation}
    \ket{\Gamma; \{j^{\alpha\beta}_{ij}\}, \{I_i\}} \label{gauge-inv-spn-ntwrk}
\end{equation}
associated with the graph $\Gamma$. Such spin network states form an orthonormal basis for the $\mathrm{SU}(2)$ gauge invariant Hilbert space $\mathcal{H}_{\Gamma}$ associated with $\Gamma$. These states can easily be constructed in the GFT pre-Fock space \eqref{gft-pre-fock}. From Section \ref{sec:gft-reps}, we know that this space contains open spin networks which have $N$ $d$-valent vertices, with the following spin and intertwiner data: the $\alpha$th link on the $i$th vertex carries a label $(j^{\alpha}_i, m^{\alpha}_i$), while the vertex itself carries an intertwiner $I_i$ between the Hilbert spaces from which the $j^{\alpha}_i$ come. We thus get an \textit{open} spin network state, namely
\begin{equation}
    \ket{\{(j^{\alpha}_i, m^{\alpha}_i)\}, \{I_i\}}\,.
\end{equation}
These states can be used to construct the $\mathrm{SU}(2)$ invariant spin network state above \eqref{gauge-inv-spn-ntwrk}, for the construction consists merely in connecting the appropriate links on distinct open spin network vertices. For instance, the labels $(j^\alpha_i, m^\alpha_i)$ and $(j^\beta_k m^\beta_k)$ on the $i$th and $k$th vertices should both coincide with the label $j^{\alpha\beta}_{ik}$ on the LQG spin network. This can be achieved by letting $(j^\alpha_i, m^\alpha_i) = (j^\beta_k, m^\beta_k)$ and summing over the magnetic quantum numbers. Therefore, more generally, we have
\begin{equation}
    \ket{\Gamma; \{j^{\alpha\beta}_{ij}\}, \{I_i\}} = \sum_{\{m^\alpha_i\}} \ket{ \{(j^{\alpha}_i, m^{\alpha}_i)\}, \{I_i\}} \prod_{(i\alpha,j\beta) \in E(\Gamma)} \delta_{j^\alpha_i, j^\beta_k} \delta_{m^\alpha_i, m^\beta_k}  \in \mathcal{H}_{\Gamma}\, , \label{eq3.46}
\end{equation}
where $E(\Gamma)$ is the set of edges in $\Gamma$. Now, using $\mathrm{SU}(2)$ recoupling theory \cite{1962mata.book.....Y,1988qtam.book.....V}, the intertwiner at every vertex in $\ket{\Gamma, \{(j^{\alpha}_i, m^{\alpha}_i)\}, \{I_i\}}$ can be repeatedly split up into products of sums of intertwiners that connect a subspace of the spins intersecting that vertex, until one obtains a spin network that has only 4-valent vertices. In other words, $\ket{\Gamma, \{(j^{\alpha}_i, m^{\alpha}_i)\}, \{I_i\}}$ can be expanded in terms of open spin networks, and hence lies in the pre-Fock space $\tilde{\mathcal{F}}_{\text{GFT}}$ that we saw above. 
 
The question of whether $\ket{\Gamma; \{j^{\alpha\beta}_{ij}\}, \{I_i\}}$ lies in the \textit{true} GFT Fock space $\mathcal{F}_{\text{GFT}}$ \eqref{gft-fock-sp} rather than just the pre-Fock space $\tilde{\mathcal{F}}_{\text{GFT}}$ \eqref{gft-pre-fock} is a bit more involved. Here we must bear in mind that the graphs $\Gamma$ under consideration so far are \textit{labeled}, i.e. their vertices are ordered by construction. On the other hand, in the GFT Fock space \eqref{gft-fock-sp}, like any other Fock space, one symmetrizes over the ``single-particle'' vertex Hilbert spaces $\mathcal{H}_v$ in defining $\mathcal{H}_V$. This Fock space, therefore, should only contain spin networks whose underlying graphs have indistinguishable vertices; call such graphs \textit{unlabeled}. From the perspective of GFT, this is not unreasonable: it is only the combinatorial pattern of vertex-edge connections that is of relevance for the geometrical information encoded in a graph \cite{Colafranceschi:2020ern}. For example, the three-vertex labeled graph $3-1-2$ should be the same as the three-vertex labeled graph $1-2-3$ or $1-3-2$, etc. (here the dashes denote edges and the number denote vertices). 
With this in mind, one should aim to show that spin networks of labeled graphs can be projected onto spin networks of unlabeled graphs. To do this, note that a labeled graph can be turned into an unlabeled graph by taking equivalence classes under permutations of vertex labels. For a given labeled graph $\Gamma$ with $V$ vertices, let $[\Gamma]$ denote its equivalence class under permutations $\pi$ of $\{1,\ldots, V\}$. Then a spin network on $\Gamma$ can be projected onto a spin network on the equivalence class $[\Gamma]$ by group averaging \cite{Colafranceschi:2020ern},
\begin{equation}
    \ket{[\Gamma]; \{j^{\alpha\beta}_{ij}\}, \{I_i\}} = \sum_{\pi \in \mathcal{P}_\Gamma} \ket{\Gamma_{\pi}; \{j^{\pi^*(\alpha\beta)}_{\pi(i)\pi(j)}\}, \{I_{\pi(i)}\}}\, , \label{gft-spin-ntwrks}
\end{equation}
where $\mathcal{P}_\Gamma$ is the group of permutations of $\{1,\ldots,V\}$. Evidently, such states are invariant under vertex relabeling and thus lie in $\mathcal{F}_{\text{GFT}}$.
\myparagraph{Diffeo-invariant states.}
  In LQG one  considers graphs embedded in a 3-manifold $\Sigma$ and therefore, the space of interest is the Hilbert space $\mathcal{H}^{\text{diff}}_\Gamma$ of diffeomorphism invariant spin networks associated with a graph $\Gamma$. That is, the diffeo-invariant LQG Hilbert space $\mathcal{H}_{\text{LQG}}$ is  
\begin{equation*}
    \mathcal{H}_{\text{LQG}} = \bigoplus_{\Gamma\in \mathcal{S}} \mathcal{H}^{\text{diff}}_{\Gamma}\, ,
\end{equation*}
where $\mathcal{S}$ is the set of all graphs embedded in $\Sigma$. We thus want to see whether equivalence classes of spin networks under diffeomorphisms can be constructed in the GFT Fock space. To answer this question, it will be pedagogically useful to probe the structure of $\mathcal{H}^{\text{diff}}_\Gamma$ in a bit more detail.

$\mathcal{H}^{\text{diff}}_\Gamma$ can be constructed from states in $\mathcal{H}_{\Gamma}$ by group averaging \cite{Ashtekar2004BackgroundIQ}. This is most easily achieved by dividing the action of the diffeomorphism group $\mathrm{Diff}_\Sigma$ on a graph into two parts. The first part is $\mathrm{GS}_\Gamma = \mathrm{Diff}_\Gamma/\mathrm{TDiff}_\Gamma$, where $\mathrm{Diff}_\Gamma \subset \mathrm{Diff}_\Sigma$ is the group of diffeomorphisms which map $\Gamma$ onto itself, whereas $\mathrm{TDiff}_\Gamma \subset \mathrm{Diff}_\Gamma$ maps $\Gamma$ onto itself while preserving the edge-vertex connectivity. That is, if the edges labeled $i$ and $j$ in $\Gamma$ are connected by an edge, they remain connected by an edge under the action of any $\xi\in \mathrm{TDiff}_\Gamma$. In other words, $\mathrm{TDiff}_\Gamma$ is essentially the automorphism group of $\Gamma$. The second part of $\mathrm{Diff}_\Sigma$ is $\mathrm{Diff}_\Sigma/\mathrm{GS}_\Gamma$, namely diffeomorphisms that merely move $\Gamma$ in $\Sigma$. Then one can define the map $\mu : \mathcal{H}_{\Gamma}\to\mathcal{H}^{\mathrm{diff}}_\Gamma$ by \footnote{As is well-known, due to the non-compactness of $\mathrm{Diff}_{\Sigma}$, one has to work in the dual of $\mathcal{H}_{\Gamma}$, but for simplicity, we ignore this technical point \cite{Ashtekar2004BackgroundIQ}.}
\begin{align}
    \mu(\ket{\Gamma; \{j^{\alpha\beta}_{ij}\}, \{I_i\}}) &\equiv \ket{\Gamma; \{j^{\alpha\beta}_{ij}\}, \{I_i\}}_\mu\nonumber\\
    &=  \frac{1}{|\mathrm{GS}_\Gamma|}\sum_{\rho \in \mathrm{Diff}_\Sigma/\mathrm{GS}_\Gamma} \sum_{\varphi\in \mathrm{GS}_\Gamma} \ket{\rho^*\varphi^*\Gamma; \{j^{\alpha\beta}_{ij}\}, \{I_i\}} \label{diff-spn-ntwrk}
\end{align}
  From the perspective of comparison with GFT, a real simplification is afforded by the remarkable result in \cite{Sahlmann:2023plc}, namely that the space $\mathcal{H}_{\text{LQG}}$ of diffeomorphism invariant spin networks actually admits a Fock space structure which is different in construction from but essentially the same in spirit as the GFT Fock space $\mathcal{F}_{\text{GFT}}$ \eqref{gft-fock-sp}. This considerably facilitates the comparison between $\mathcal{F}_{\text{GFT}}$ and $\mathcal{H}_{\text{LQG}}$. We will thus briefly sketch the main ideas involved in exhibiting a Fock structure in $\mathcal{H}_{\text{LQG}}$; for details, see \cite{Sahlmann:2023plc}.

The central idea is to partition the set $\mathcal{S}$ of graphs in $\Sigma$ into distinct numbers of \textit{unlinked components}, where a single unlinked component of a graph $\Gamma$ means one or several parts of $\Gamma$ that are not linked to the rest of the graph, but are linked to each other. Let $\mathcal{S}_n$ denote the set of all graphs with $n$ unlinked components. Since the Ashtekar-Lewandowski measure ensures that the inner product between two graphs $\Gamma\in\mathcal{S}_n$ and $\Gamma'\in\mathcal{S}_m$ is zero unless $m=n$, one has
\begin{equation*}
    \mathcal{H}_{\text{LQG}} = \bigoplus_{\Gamma\in \mathcal{S}} \mathcal{H}^{\text{diff}}_{\Gamma} = \bigoplus_{n=0}^{\infty}\bigoplus_{\Gamma\in\mathcal{S}_n}\mathcal{H}^{\mathrm{diff}}_\Gamma = \bigoplus_{n=0}^{\infty} \mathcal{H}^{\mathrm{diff}}_n\, ,
\end{equation*}
where we have defined an $n$-component Hilbert space $\mathcal{H}^{\mathrm{diff}}_n := \bigoplus_{\Gamma\in\mathcal{S}_n}\mathcal{H}^{\mathrm{diff}}_\Gamma$. It can be shown \cite{Sahlmann:2023plc} that this $n$-component Hilbert space is isomorphic to $\mathrm{sym}(\mathcal{H}^{\mathrm{diff}}_1\otimes\cdots\otimes\mathcal{H}^{\mathrm{diff}}_1)$. Thus $\mathcal{H}_{\mathrm{LQG}}$ is a Fock space in which spin networks coming from one-component diffeomorphic graphs are one-particle states. Explicitly, let $\gamma_i$, $i=1,\ldots,n$, denote $n$ one-component graphs; then $n$-particle states in $\mathcal{H}_{\mathrm{LQG}}$ are spanned by
\begin{equation}
    \sum_{\sigma\in \mathcal{P}_n}\bigotimes_{i=1}^{n}\ket{\gamma_{\sigma(i)}; \{j^{\alpha\beta}_{ij}\}, \{I_i\}}_\mu\, ,
\end{equation}
where $\mathcal{P}_n$ is the symmetric group over $n$ elements, denoting the permutations between distinct components $\gamma_i$. Now, for a one-component graph $\gamma$, the sum over $\mathrm{GS}_{\gamma}$ in \eqref{diff-spn-ntwrk} essentially amounts to a sum over all permutations of the vertices of $\gamma$. This produces nothing but the associated unlabeled graph $[\gamma]$, resulting in a state in $\mathcal{F}_{\mathrm{GFT}}$. Furthermore, from the standpoint of GFT, the sum over $\mathrm{Diff}_\Sigma/\mathrm{GS}_\Gamma$ in \eqref{diff-spn-ntwrk} is redundant, since there is no embedding of graphs in an a priori manifold in GFT. In other words, moving a graph in $\Sigma$ gives exactly the same graph from a GFT perspective. Therefore, a spin network in $\mathcal{H}_{\mathrm{LQG}}$ can be constructed in $\mathcal{F}_{\mathrm{GFT}}$ as well \cite{Sahlmann:2023plc}. 

Thus, the set of all kinematical LQG states is part of the set of states in GFT Fock space $\mathcal{F}_{\text{GFT}}$. Notice, however, that despite this pleasant result, the spaces $\mathcal{H}_{\mathrm{LQG}}$ and $\mathcal{F}_{\mathrm{GFT}}$ are fundamentally different. To begin with, one-particle states in $\mathcal{H}_{\mathrm{LQG}}$ are spin networks associated with one-component graphs, whereas one-particle states in $\mathcal{F}_{\mathrm{GFT}}$ are open spin networks associated with vertices. This in turn affects how orthogonality between states is imposed in the two spaces. In $\mathcal{H}_{\mathrm{LQG}}$, two states are orthogonal either if they have a different number of components or any two components in them are not diffeomorphic to each other \cite{Sahlmann:2023plc}. On the other hand, states in $\mathcal{F}_{\mathrm{GFT}}$ are orthogonal if they have a different number of vertices. In particular, we may consider two states coming, respectively, from a graph $\Gamma$ with two unlinked components each of which has, say, two vertices, and a graph $\Gamma'$ with only one component which has, say, four vertices. In $\mathcal{H}_{\mathrm{LQG}}$, their inner product will vanish, but in $\mathcal{F}_{\mathrm{GFT}}$, they may have a nonzero overlap. This is not a paradox. Graphs with the same number of vertices but different number of unlinked components contain an equal number of quanta of spacetime. They may differ only with respect to their combinatorial patterns, but in the absence of a continuous manifold with fixed topology in which one embeds graphs as in LQG, there is no reason to suppose that the Hilbert space of quantum geometry would partition into sectors with distinct combinatorial patterns \cite{Colafranceschi:2020ern}. The only point in common in the two approaches at this level of abstraction is the fact that graphs with a distinct number of vertices encode different geometries (areas, volumes, etc.). Consequently, their overlap should be zero, and this is indeed what happens both in $\mathcal{H}_{\mathrm{LQG}}$ and $\mathcal{F}_{\mathrm{GFT}}$. 

\subsection{Towards a GFT Fock space for the HK model}\label{sec:4.2}
In view of the differences between $\mathcal{H}_{\mathrm{LQG}}$ and $\mathcal{F}_{\mathrm{GFT}}$, the GFT Fock space stipulated above should be regarded as establishing a connection with a notion of canonical quantization somewhat different, and perhaps more general, than that of canonical LQG. That we are justified in doing so, at least within the context of the HK model, emerges from the following considerations. Fundamentally, the canonical quantization of the HK model consists in defining precisely the formula \cite{Husain:1990sc}
\begin{equation}
    \braket{\psi, \psi'} = \int 
    DA\,\Bar{\psi}(A)\psi'(A) \sim \delta_{\psi,\psi'} \label{informal-inn-prd}\, .
\end{equation}
Here $\psi, \psi'$ are some states in a suitable physical Hilbert space representing distinct geometries, and $DA$ 
denotes a measure over the space of all connections $\mathrm{SU(2)}$ $A$ on a manifold. As we have seen above, in $\mathcal{F}_{\mathrm{GFT}}$, states with a distinct number of vertices encode a distinct geometry, and accordingly, states with a different number of vertices have zero overlap, whereas graphs with the same number of vertices do not. This is in conformity with \eqref{informal-inn-prd}.  

Incidentally, the preceding remarks suggest that the GFT Fock space considered above should be the physical Hilbert space for a GFT of the HK model. But we mentioned neither the HK model nor any other theory of gravity in constructing (or rather, postulating) those Fock spaces. So how do we know that we  have a quantum description of the HK model? And if we have, how is it different from the Fock space for a GFT for a different classical model? 

To bring these questions into relief, let us inquire into what it means to ``quantize'' a GFT. One possible meaning is furnished by the partition function of GFT and another by building a Fock space from GFT fields as above. In usual field theories, these two paths to quantization are in principle independent, and possibly in general inequivalent. To elaborate, contrast the situation here with, for instance, scalar field theory on Minkowski spacetime. There, the Fock space vacuum is fixed by its being a zero energy eigenstate of the (normal-ordered) Hamiltonian. On the other hand, in the path integral quantization, one can define a vacuum indirectly by using correlation functions, which in turn depend on the Green function of the theory \cite{Weinberg:1995mt}. But in constructing a GFT Fock space, we made no reference to a Hamiltonian; the construction is oblivious to the content of the interaction term in the GFT action. But evidently, different actions should yield different correlation functions, which presumably correspond to different vacua. Which of these correspond to the vacuum above? In Section \ref{sec:fock}, we provide an answer to this question by using the methods of algebraic quantum field theory to show that the Fock space that was considered above is the Fock space of a non-interacting GFT (no interaction term in the GFT action), and that this, suitably interpreted, is the GFT of the HK model. 

Thus, at least for the HK model, we can show the explicit connection between the path integral and Fock space quantization of the corresponding GFT. In turn, this shows that states in $\mathcal{F}_{\mathrm{GFT}}$ do in fact enable one to rigorously define the canonical inner product \eqref{informal-inn-prd} for the HK model. 

\section{GFT quantization of the HK model}\label{sec:hk-gft}
As we have seen in the review in Sections \ref{sec:intro-gft} and \ref{sec:gft-fock}, GFTs provide a formalism to possibly serve as a bridge between canonical quantization and spinfoams. In this section, we will see this concretely in the case of the HK model. We will show this by first constructing a GFT model that provides a completion of the HK spinfoam model of \cite{Bojowald:2009im}. This will be done in Section \ref{sec:pathintegral}. Then, in Section \ref{sec:fock}, we will show that a unique Fock representation is associated with this GFT model. This Fock space will turn out to be a subspace of the Fock space considered in the previous section. Thus, since amplitudes assigned in this Fock construction to arbitrary abstract spin-network states have already been shown to match the ones expected from an LQG-inspired canonical quantization of the HK model, we will have provided an explicit connection between the spinfoam and LQG quantizations of this model. 

\subsection{Path-integral representation }\label{sec:pathintegral}
\begin{figure}[ht]
    \centering
    \begin{tikzpicture}[x=0.75pt,y=0.75pt,yscale=-1,xscale=1]
        \draw   (153,70) -- (206.26,108.7) -- (185.92,171.3) -- (120.08,171.3) -- (99.74,108.7) -- cycle ;
        
        \node[above] at (153,72) {$\psi$};
        \node[right] at (206.26,108.7) {$\psi$};
        \node[right] at (185.92,171.3) {$\psi$};
        \node[left] at (120.08,171.3) {$\phi$};
        \node[left] at (99.74,108.7) {$\phi$};
        
        \node[above] at (178,128) {$j_4$};
        \node[above] at (130,128) {$j_3$};
        \node[above] at (153,108.7) {$j_{10}$};
        \node[below] at (170,140) {$j_5$};
        \node[below] at (135,142) {$j_9$};
        
        \node[above right] at (193,150) {$j_6$};
        \node[above left] at (113,150) {$j_8$};
        \node[below] at (186,75) {$j_2$};
        \node[below right] at (148,171.3) {$j_7$};
        \node[below left] at (130,75) {$j_1$};
        
        \draw    (99.74,108.7) -- (185.92,171.3) ;
        \draw    (206.26,108.7) -- (120.08,171.3) ;
        \draw    (99.74,108.7) -- (206.26,108.7) ;
        \draw    (153,70) -- (185.92,171.3) ;
        \draw    (153,70) -- (120.08,171.3) ;

\draw    (232,120) -- (301,120.97) ;
\draw [shift={(303,121)}, rotate = 180.81] [color={rgb, 255:red, 0; green, 0; blue, 0 }  ][line width=0.75]    (10.93,-3.29) .. controls (6.95,-1.4) and (3.31,-0.3) .. (0,0) .. controls (3.31,0.3) and (6.95,1.4) .. (10.93,3.29)   ;

\draw    (328.6,163.14) .. controls (351.72,178) and (432.45,178.79) .. (462.77,155.32) ;
\draw [shift={(462.77,155.32)}, rotate = 322.26] [color={rgb, 255:red, 0; green, 0; blue, 0 }  ][fill={rgb, 255:red, 0; green, 0; blue, 0 }  ][line width=0.75]      (0, 0) circle [x radius= 3.35, y radius= 3.35]   ;
\draw    (328.6,163.14) .. controls (325.57,124.8) and (314.2,52.83) .. (373.32,85.69) ;
\draw [shift={(328.6,163.14)}, rotate = 265.48] [color={rgb, 255:red, 0; green, 0; blue, 0 }  ][fill={rgb, 255:red, 0; green, 0; blue, 0 }  ][line width=0.75]      (0, 0) circle [x radius= 3.35, y radius= 3.35]   ;
\draw    (328.6,163.14) .. controls (344.52,122.46) and (405.16,35.62) .. (437.76,77.08) ;
\draw    (383.18,91.16) .. controls (418.81,109.16) and (454.43,141.23) .. (462.77,155.32) ;
\draw    (328.6,163.14) .. controls (346.79,131.85) and (373.32,111.5) .. (393.03,103.68) ;
\draw    (403.64,98.99) .. controls (502.95,47.35) and (484,105.25) .. (462.77,155.32) ;
\draw    (462.77,155.32) .. controls (458.98,121.68) and (449.13,103.68) .. (442.3,87.25) ;

\draw (387.81,175.66) node [anchor=north west][inner sep=0.75pt]    {$j_{8}$};
\draw (477.25,68.48) node [anchor=north west][inner sep=0.75pt]    {$j_{5}$};
\draw (407.51,45.79) node [anchor=north west][inner sep=0.75pt]    {$j_{9}$};
\draw (321.86,63) node [anchor=north west][inner sep=0.75pt]    {$j_{1}$};
\draw (313,164.15) node [anchor=north west][inner sep=0.75pt]    {$\phi $};
\draw (463.85,152.41) node [anchor=north west][inner sep=0.75pt]    {$\phi $};

\end{tikzpicture}

    \caption{In the classical HK model, there exists a preferred direction that is perpendicular to hypersurfaces of a foliation of spacetime which are spanned by the triads. As shown in \cite{Bojowald:2009im}, this fact can be translated to a restriction on the vertex amplitude of a $\mathrm{BF}$ theory spinfoam.} 
    \label{fig1}
\end{figure}
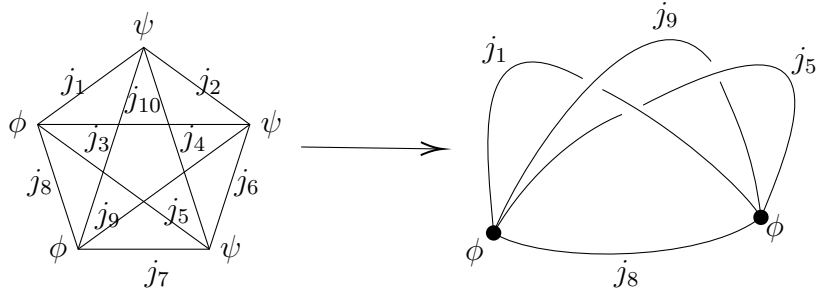

Given the physical interpretation of the HK model as a theory of non-dynamical 3-geometries and the fact that the GFT field represents configurations of ``atoms'' of such $3$-geometries, it is natural to expect that a GFT quantization of the HK model will lead to a non-interacting GFT, i.e., defined by an action with only a kinetic term and no interacting terms. In Section \ref{sec:gft-amps}, we have observed that GFTs can, in principle, be systematically constructed by determining the form of simplicial gravity path-integrals or spinfoam amplitudes. Therefore, to confirm our expectation that a GFT quantization of the HK model would result in a non-interacting theory, it is natural first to investigate the form that a spinfoam quantization of the HK model would take.
\myparagraph{The HK spinfoam vertex amplitude.} This was done in \cite{Bojowald:2009im}, by imposing constraints on an $\mathrm{SU(2)}$ $\mathrm{BF}$ theory spinfoam vertex amplitude; the argument is based on observing that in the HK model, the 2-form $B=e\wedge e$ is orthogonal to the ``special'' direction $\Tilde{u}^\alpha$ (\ref{ut}) (Section \ref{sec:hk}). We use the same fact to establish the same result (Fig. \ref{fig1}), but we state the argument a little differently. 

In a discretized $\mathrm{BF}$ theory on the triangulation of a manifold (Appendix \ref{sec:appC}), the $B$ fields in every tetrahedron of the triangulation are smeared over the faces of the tetrahedron \cite{Engle:2007qf}. The field $\tilde{u}$ can be discretized over the vertices of the dual 2-complex of the triangulation. Hence, every 4-simplex of the triangulation has a fixed value of $\tilde{u}$ associated to it. This fixed direction associated to every 4-simplex divides the tetrahedra in the 4-simplex into two types that we call ``orthogonal'' and ``non-orthogonal''. These are defined as follows. 

Consider the dual graph of a 4-simplex, shown in the left part of Fig. \ref{fig1}. Every triangle in the graph defines a unique direction (its surface normal) in the underlying 4d manifold. Furthermore, every triangle is shared among a set of three tetrahedra. For example, the triangle labeled by $(j_2,j_4,j_6)$ is shared by the three tetrahedra labeled by $\psi$. Suppose now that this triangle is orthogonal to $\tilde{u}$. Then we call the tetrahedra labeled by $\psi$ the orthogonal tetrahedra, whereas the remaining tetrahedra (labeled by $\phi$) are called non-orthogonal. 

Now, upon quantization, the smeared $B$ fields are associated with angular momentum operators $J^i$ (three for three internal components of the $\mathfrak{su}(2)$-valued $B$ field), and so, passing to the spin representation, every face of a tetrahedron carries a spin label. For 4-dimensional $\mathrm{BF}$ theory, the vertex amplitude is represented by the dual graph of a 4-simplex, the edges carrying the spin labels and the vertices carrying intertwiners \cite{Baez:1999sr}, as shown in the left of Fig. \ref{fig1}. A plausible way to interpret the constraint $B^i_{\alpha\beta}\tilde{u}^\alpha = 0$ is to set the spins on the faces shared by the orthogonal tetrahedra to zero \cite{Bojowald:2009im}. For instance, in Fig. \ref{fig1}, one would set $j_2=j_4=j_6=0$; this reduces the 5-vertex graph dual to a 4-simplex to the 2-vertex graph shown in the right. Thus, in the full spinfoam amplitude, which contains a product over all 4-simplices in the triangulation (Appendix \ref{sec:appC}), one would set the amplitude at every 4-simplex to zero unless at least one triangle in its dual graph has all its edges labeled by zero spin. Since the amplitude also contains a product over all faces dual to the triangles in the underlying triangulation, all triangles are ``integrated out'' in this way, and hence all directions of $\tilde{u}$ are accounted for, and the resulting spinfoam amplitude will be independent of $\tilde{u}$, as it should be \cite{Bojowald:2009im}. In short, the full amplitude will reduce to a sum over 2-vertex spin networks shown in Fig. \ref{fig1}. Schematically, 
\begin{equation}
    Z_{\text{HK}}(C) = \sum_{j}\prod_{f^*\in C^*}(2j_{f^*}+1)\prod_{v\in\mathcal{V}}\{4j\}~,
\end{equation}
where $C$ is the triangulation of the underlying manifold and $C^*$ its dual 2-complex with the set of vertices $\mathcal{V}$, and $\{4j\}$ symbolically stands for the spin-network evaluation of the graph in the right part of Fig. \ref{fig1}. 

\myparagraph{A GFT action for the HK model.}
The spinfoam amplitude for the HK model derived above paves the way for writing down a GFT action for the HK model. We now describe how this can be done. 

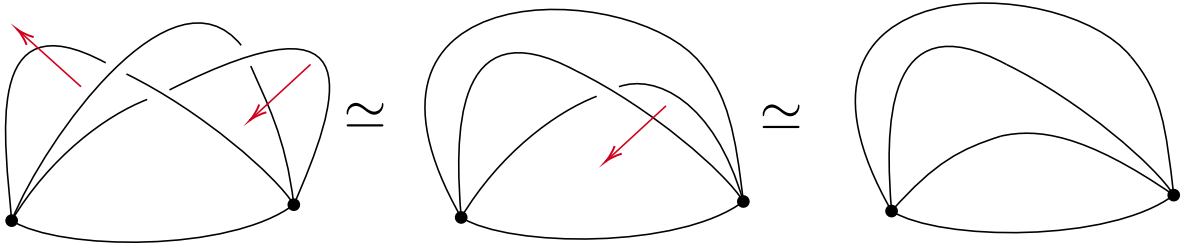
\begin{figure}[ht]
    \centering

\tikzset{every picture/.style={line width=0.6pt}} 

\begin{tikzpicture}[x=0.6pt,y=0.6pt,yscale=-1,xscale=1]

\draw    (47.5,314) .. controls (78,333) and (184.5,334) .. (224.5,304) ;
\draw [shift={(224.5,304)}, rotate = 323.13] [color={rgb, 255:red, 0; green, 0; blue, 0 }  ][fill={rgb, 255:red, 0; green, 0; blue, 0 }  ][line width=0.75]      (0, 0) circle [x radius= 3.35, y radius= 3.35]   ;
\draw    (47.5,314) .. controls (43.5,265) and (28.5,173) .. (106.5,215) ;
\draw [shift={(47.5,314)}, rotate = 265.33] [color={rgb, 255:red, 0; green, 0; blue, 0 }  ][fill={rgb, 255:red, 0; green, 0; blue, 0 }  ][line width=0.75]      (0, 0) circle [x radius= 3.35, y radius= 3.35]   ;
\draw    (47.5,314) .. controls (68.5,262) and (148.5,151) .. (191.5,204) ;
\draw    (119.5,222) .. controls (166.5,245) and (213.5,286) .. (224.5,304) ;
\draw    (47.5,314) .. controls (71.5,274) and (106.5,248) .. (132.5,238) ;
\draw    (146.5,232) .. controls (277.5,166) and (252.5,240) .. (224.5,304) ;
\draw    (224.5,304) .. controls (219.5,261) and (206.5,238) .. (197.5,217) ;

\draw    (329.5,312) .. controls (360,331) and (466.5,332) .. (506.5,302) ;
\draw [shift={(506.5,302)}, rotate = 323.13] [color={rgb, 255:red, 0; green, 0; blue, 0 }  ][fill={rgb, 255:red, 0; green, 0; blue, 0 }  ][line width=0.75]      (0, 0) circle [x radius= 3.35, y radius= 3.35]   ;
\draw    (329.5,312) .. controls (325.5,263) and (323.5,178) .. (401.5,220) ;
\draw [shift={(329.5,312)}, rotate = 265.33] [color={rgb, 255:red, 0; green, 0; blue, 0 }  ][fill={rgb, 255:red, 0; green, 0; blue, 0 }  ][line width=0.75]      (0, 0) circle [x radius= 3.35, y radius= 3.35]   ;
\draw    (329.5,312) .. controls (241,157) and (436.5,162) .. (479.5,215) ;
\draw    (401.5,220) .. controls (448.5,243) and (495.5,284) .. (506.5,302) ;
\draw    (329.5,312) .. controls (353.5,272) and (388.5,246) .. (414.5,236) ;
\draw    (428.5,230) .. controls (451,222) and (488,238) .. (506.5,302) ;
\draw    (506.5,302) .. controls (501.5,259) and (497,236) .. (479.5,215) ;
\draw    (599.5,308) .. controls (630,327) and (736.5,328) .. (776.5,298) ;
\draw [shift={(776.5,298)}, rotate = 323.13] [color={rgb, 255:red, 0; green, 0; blue, 0 }  ][fill={rgb, 255:red, 0; green, 0; blue, 0 }  ][line width=0.75]      (0, 0) circle [x radius= 3.35, y radius= 3.35]   ;
\draw    (599.5,308) .. controls (595.5,259) and (593.5,174) .. (671.5,216) ;
\draw [shift={(599.5,308)}, rotate = 265.33] [color={rgb, 255:red, 0; green, 0; blue, 0 }  ][fill={rgb, 255:red, 0; green, 0; blue, 0 }  ][line width=0.75]      (0, 0) circle [x radius= 3.35, y radius= 3.35]   ;
\draw    (599.5,308) .. controls (511,153) and (706.5,158) .. (749.5,211) ;
\draw    (671.5,216) .. controls (718.5,239) and (765.5,280) .. (776.5,298) ;
\draw    (599.5,308) .. controls (627,277) and (643,269) .. (668,261) ;
\draw    (668,261) .. controls (690.5,256) and (720,259) .. (776.5,298) ;
\draw    (776.5,298) .. controls (771.5,255) and (769,234) .. (749.5,211) ;
\draw [color={rgb, 255:red, 208; green, 2; blue, 27 }  ,draw opacity=1 ]   (235,216) -- (198.47,249.65) ;
\draw [shift={(197,251)}, rotate = 317.35] [color={rgb, 255:red, 208; green, 2; blue, 27 }  ,draw opacity=1 ][line width=0.75]    (10.93,-3.29) .. controls (6.95,-1.4) and (3.31,-0.3) .. (0,0) .. controls (3.31,0.3) and (6.95,1.4) .. (10.93,3.29)   ;
\draw [color={rgb, 255:red, 208; green, 2; blue, 27 }  ,draw opacity=1 ]   (91,230) -- (52.49,195.34) ;
\draw [shift={(51,194)}, rotate = 41.99] [color={rgb, 255:red, 208; green, 2; blue, 27 }  ,draw opacity=1 ][line width=0.75]    (10.93,-3.29) .. controls (6.95,-1.4) and (3.31,-0.3) .. (0,0) .. controls (3.31,0.3) and (6.95,1.4) .. (10.93,3.29)   ;
\draw [color={rgb, 255:red, 208; green, 2; blue, 27 }  ,draw opacity=1 ]   (458,242) -- (421.47,275.65) ;
\draw [shift={(420,277)}, rotate = 317.35] [color={rgb, 255:red, 208; green, 2; blue, 27 }  ,draw opacity=1 ][line width=0.75]    (10.93,-3.29) .. controls (6.95,-1.4) and (3.31,-0.3) .. (0,0) .. controls (3.31,0.3) and (6.95,1.4) .. (10.93,3.29)   ;

\draw (254,238.4) node [anchor=north west][inner sep=0.75pt]  [font=\LARGE]  {${\displaystyle \simeq }$};
\draw (515,239.4) node [anchor=north west][inner sep=0.75pt]  [font=\LARGE]  {${\displaystyle \simeq }$};

\end{tikzpicture}
    \caption{Planar isotopy of the 2-vertex graph in Fig. \ref{fig1}. The red arrow(s) in a diagram indicates the type II Reidemaster move(s) that can be performed to obtain the subsequent diagram.}
    \label{fig:planar-iso}
\end{figure}

As we have seen, the HK model, despite being formulated on a 4d manifold, is effectively a theory of 3-geometries. This is evident also at the quantum level by looking at the restricted vertex amplitude 
presented above: the  graph in the right part of Fig.~\ref{fig1} is dual to the triangulation of a 3-sphere
, as it is planar isotopic to a standard melonic graph (see Fig. \ref{fig:planar-iso}). The standard melon, in turn, 
is the dual graph of two tetrahedra glued together to form a triangulation of a 3-sphere. 

From the above argument, we deduce that the physics of the quantum HK model encoded in the above amplitude, is in fact 
agnostic about which link on one vertex is connected with which link on the second vertex. This is unlike the case in the full unrestricted $\mathrm{BF}$-theory amplitude, where the full graph dual to a 4-simplex contains information about the precise way in which the faces of all the tetrahedra in the 4-simplex are glued in order to produce a 4-simplex \cite{Gurau:2009tw}. This makes sense because the combinatorial structures in a 4d spinfoam theory have to describe triangulations of 4d manifolds. In contrast, the HK model describes 3-geometries because, as we have argued above, its vertex graph is dual to two tetrahedra glued together to triangulate a 3-sphere. Now, the order in which the faces of the two tetrahedra are glued does not matter; one will get a triangulation of a 3-sphere all the same. 

How is this all translated to a GFT context?  Recall that a GFT ``interaction term'' describing a melon should be of the form $\int \diff g_1\cdots \diff g_4 ~\phi_{1234}\phi_{1234}$. However, as it stands, such a term actually describes something more: it corresponds to a stranded graph composed of two 4-valent vertices in which the order in which the links on the two vertices are connected is fixed (see Fig. \ref{fig:vertex-prop}a). Such a stranded graph can be thought of as describing a melon, but so can the stranded graph corresponding to a quadratic term with any permutation of $(1234)$ in one of the $\phi$'s, e.g. $\phi_{1234}\phi_{1324}$. 
In order to effectively avoid this redundancy,  
one can demand that the group field $\phi$ be invariant under all permutations of its arguments.

To make the above discussion more concrete, we can mimic the kind of vertex restriction of a spinfoam amplitude presented above in a GFT setting. Recall that the spinfoam amplitude for the HK model was derived by restricting the vertex amplitude of a four-dimensional $\mathrm{BF}$ theory. From a GFT perspective, a four-dimensional $\mathrm{BF}$ theory corresponds to the Ooguri model that we presented in Section \ref{sec:gft-amps}. 
Thus, it is natural to construct a GFT quantization of the HK model by following an effective vertex reduction at the level of GFT similar to the one performed in the spinfoam theory, i.e.\ by considering a full 4d GFT action corresponding to four-dimensional $\mathrm{BF}$ theory, namely the Ooguri model \eqref{Oaction}, and imposing on the action an appropriate constraint that produces something like Fig. \ref{fig1}. 

This can be done by using two group fields $\phi$ and $\psi$, with one of them appearing in the constraint term. As anticipated in the discussion of the spinfoam theory, one could think of the above two fields as generating tetrahedra that are orthogonal ($\psi$) or not ($\phi$) with respect to the preferred HK direction\footnote{This could be made explicit by extending the GFT field domain to include a normal vector, as done, e.g.\ in \cite{Baratin:2011tx,Jercher:2022mky}.}. Bearing this in mind, we start with the action
\begin{align}
    S_C &= \frac{1}{2}\int \diff g_1\cdots \diff g_4\ \phi^2(g_1, g_2, g_3, g_4) \nonumber \\ 
    &- \frac{\lambda}{2}\int \diff g_1\cdots \diff g_{10}\ \psi(g_1, g_2, g_3, g_4)\psi(g_4, g_5, g_6, g_7)\psi(g_9, g_6, g_2, g_{10}) \nonumber\\
&\qquad\qquad\qquad\times\phi(g_7, g_3, g_8, g_9)\phi(g_{10}, g_8, g_5, g_1) \, , \label{eq8}
\end{align}
where the field $\psi$ is subject to the constraint 
\begin{align}
    \int \diff g_2\diff g_4\diff g_6\, &\psi(g_1, g_2, g_3, g_4)\psi(g_4, g_5, g_6, g_7)\psi(g_9, g_6, g_2, g_{10})\nonumber \\ 
    &- \delta(g_9^{-1}g_{10})\delta(g_1^{-1}g_3)\delta(g_5^{-1}g_7) = 0\, . \label{gft-cons} 
\end{align}
This constraint yields a restriction of the GFT interaction equivalent to the spinfoam vertex restriction of \cite{Bojowald:2009im} described above in Fig. \ref{fig1}, further confirming the physical interpretation provided above for the field\footnote{Our aim here is really to just consider $\psi$ as a field with trivial (i.e. constrained) dynamics, i.e., just as a field allowing us to implement the existence of a preferred direction in HK consistently at the quantum level. One could, in principle, attempt to include a kinetic term for $\psi$ as well, making it a \qmarks{dynamical} field alongside $\phi$. However, this would go beyond the scope of this paper and might lead to a model that corresponds not to the quantization of exact HK but rather to a slightly modified version of it.} $\psi$. Indeed, let us substitute \eqref{spinrep1} into \eqref{gft-cons}. We get
\begin{align*}
    \int \diff g_2 \diff g_4 \diff g_6&\sum_{JMNN'j}\sum_{KPQQ'k}\sum_{LRSS'l}\sqrt{d_Jd_Kd_L}\psi^{J}_{MN'}\psi^{K}_{PQ'}\psi^{L}_{RS'}I^{Jj}_{N'}I^{Jj}_{N}I^{Kk}_{Q'}I^{Kk}_{Q}I^{Ll}_{S'}I^{Ll}_{S}\\
   &\times D^{j_1}_{m_1n_1}(g_1)D^{j_2}_{m_2n_2}(g_2)D^{j_3}_{m_3n_3}(g_3)D^{j_4}_{m_4n_4}(g_4)\\
&\times D^{k_1}_{p_1q_1}(g_4)D^{k_2}_{p_2q_2}(g_5)D^{k_3}_{p_3q_3}(g_6)D^{k_4}_{p_4q_4}(g_7)\\
&\times D^{l_1}_{r_1s_1}(g_9)D^{l_2}_{r_2s_2}(g_6)D^{l_3}_{r_3s_3}(g_2)D^{l_4}_{r_4s_4}(g_{10}) \\
&= \sum_{jklmnpqrs}d_jd_kd_l(-1)^{j+k+l-(m+p+r)}D^{j}_{-m-n}(g_9)D^j_{mn}(g_{10})D^{k}_{-p-q}(g_1)D^k_{pq}(g_3)\\
&\qquad\qquad\qquad\qquad\qquad\qquad\qquad\qquad\times D^{l}_{-r-s}(g_5)D^l_{rs}(g_5)\,.
\end{align*}
Using the identities \eqref{orth-1}-\eqref{peter-weyl-delta}, it is straightforward to see that in order for the preceding equation to be true, $j_1=j_3$, $k_2=k_4$, $l_1=l_4$, $k_1=j_4$, $l_2=k_3$, $l_3=j_2$ and $j_2=j_4=k=3=0$. This is exactly what is shown in Fig. \ref{fig1}. 

Eqn. \eqref{gft-cons} is only one possible constraint term. In general, there will be 9 more, corresponding to the 9 triangles in the left of Fig. \ref{fig1} other than the one labeled by $(j_2,j_4,j_6)$ considered above. If one substitutes \eqref{gft-cons} into $S_C$ above, one obtains
\begin{equation}
    S = \frac{1}{2}\int \diff g_1\cdots \diff g_4~ \phi^2(g_1,\ldots,g_4) - \frac{\lambda}{2}\int \diff g_1\cdots \diff g_4 ~\phi(g_1, g_2, g_3, g_4)\phi(g_3, g_4, g_1, g_2)~.
\end{equation}
Incorporating the other constraints corresponding to the 9 other triangles, we would get combinations different from $(g_3,g_4,g_1,g_2)$ in the second term above. But all of these terms are supposed to describe the same physics, as we argued above. Thus, it is natural to demand that $\phi$ be invariant under all permutations of its arguments. Then all interaction terms are equivalent to the standard $\phi_{1234}\phi_{1234}$. In effect, we may set $\lambda = 0$ and write the GFT action for the HK model as
\begin{equation}
    S_{\text{HK-GFT}} = \frac{1}{2}\int \diff g_I\, \phi^2(g_I)\,. \label{action-hk-gft}
\end{equation}
This is the action that we shall study from now on. 

\myparagraph{Correlation functions.}
Eqn. \eqref{action-hk-gft} describes non-interacting GFT. Correlation functions for a theory can be immediately calculated by defining (we are again abbreviating $(g_1,\ldots,g_4)=(g_I)$)
\begin{equation}
    Z[J] = \int D\phi \exp{\left(-\frac{1}{2}\int \diff g_I\,\phi^2(g_I) + \int \diff g_I\,J(g_I)\phi(g_I) \right)}\, ; \label{pf-hk}
\end{equation}
we have 
\begin{align}
    \braket{\hat{\phi}(g_I)\hat{\phi}(h_I)} &= \frac{1}{Z[0]}\left.\frac{\delta}{\delta J(g_I)}\frac{\delta}{\delta J(h_I)}Z[J]\right\vert_{J=0} \label{2-pt-fn-hk}\\
    &= K^{-1}(g_I,h_I)\, ,
\end{align}
where $K^{-1}(g_I,h_I)$ is the Green function for the kernel of $\int dg_I\,\phi^2(g_I)$, i.e. the propagator of the theory. Taking into account the right-invariance of the fields, it is given by (cf. \eqref{free-prop})
\begin{equation}
    K^{-1}(g_I,h_I) =  \int \diff s\, \prod_{i=1}^4\delta(g_ish^{-1}_i) \label{prop-hk}\,.
\end{equation}
More generally, using the Wick theorem, we can also write $n$-point correlation functions in terms of the Green function:
\begin{equation}\label{eqn:npointfunctions}
    \braket{\phi(g_{I_1})\cdots\phi(g_{I_n})} = \sum_{\text{all pairings}}\prod_{i\neq j}^{n}K^{-1}(g_{I_i}, g_{I_j})\, . 
\end{equation}
As we shall see in the following, the above equation can also be obtained within an operator formalism. Furthermore, the operators in question will be represented as operators acting on a Fock space. This Fock space will turn out to be the same as a subspace of the Fock space postulated in Section \ref{sec:fock}. As we have seen, amplitudes between physical boundary states are non-zero (and in fact, trivial) only for identical boundary states in that Fock space. This, therefore, is the information implicitly contained in the preceding equation, and as such, the equation realizes the fundamental requirement of a quantization of the HK model. 

\subsection{Hilbert space representation of HK GFT}\label{sec:fock}
Given that the physical content of the above GFT model is entirely encoded in \eqref{eqn:npointfunctions}, one could attempt to define an operator theory that satisfies \eqref{eqn:npointfunctions} by constructing  expectation values of field operators $\hat{\phi}(g_I)$ in a vacuum $\ket{0}$.

Can such an operator theory be given a Hilbert space representation? To put this question in perspective, let us recall how a Hilbert space is constructed in standard quantum field theory \cite{Wald:1995yp, Ashtekar-magnon}. To be concrete, consider a free, real Klein-Gordon scalar field on a spacetime manifold. One begins with the phase space of the theory, which can be regarded as the set of initial data for the Klein-Gordon equation. The phase space comes equipped with a symplectic structure $\Omega$ that may be used to define a one-particle Hilbert space $\mathcal{H}$ and a Fock space. One first picks a space of functions $\mathcal{S}$ suitable enough to characterize the phase space (e.g. Schwartz space). Since the quantum mechanics of a single particle requires the use of a complex Hilbert space, we first complexify $\mathcal{S}$. This can be done by putting a complex structure $J$ on $\mathcal{S}$; then $\mathcal{S}_{\mathbb{C}} = \mathcal{S}+i\mathcal{S}$. The completion of $\mathcal{S}_{\mathbb{C}}$ in a suitable inner product $(\cdot,\cdot)$ gives the sought-after Hilbert space $\mathcal{H}$. The inner product in question is to be picked based on the requirement that we want it to be such that classical observables can be realized as operators on a Fock space, and that the algebra of these operators respects the classical symplectic structure. These requirements can be used to specify $(\cdot,\cdot)$ in terms of $J$ and $\Omega$ as follows. 

Suppose that we are given an inner product $(\cdot,\cdot)$ and hence $\mathcal{H}$. We define a symmetric Fock space based on $\mathcal{H}$, i.e.
\begin{equation}
    \mathcal{F}(\mathcal{H}) = \bigoplus_{n=0}^{\infty}\text{sym}\left(\mathcal{H}\underbrace{\otimes\cdots\otimes}_{n \text{ times }}\mathcal{H}\right)\,. \label{AM-comms}
\end{equation}
Quantum observables can be realized as operator-valued distributions, which are continuous linear functionals from $\mathcal{H}$ to operators on $\mathcal{F}(\mathcal{H})$. Of particular interest are the creation and annihilation operator-valued distributions $\hat{A}^\dagger$ and $\hat{A}$, which are defined by the condition that they satisfy the following algebra,
\begin{equation*}
    [\hat{A}(f_1),\hat{A}(f_2)] = [\hat{A}^\dagger(f_1), \hat{A}^\dagger(f_2)] = 0, \quad [\hat{A}(f_1), \hat{A}^\dagger(f_2)] = (f_1, f_2) \quad \forall f_1,f_2\in\mathcal{H}.
\end{equation*}
Since, for every $f\in\mathcal{H}$, the operator corresponding to the Klein-Gordon field is given by $\hat{\phi}(f)=\hat{A}(f)+\hat{A}^\dagger(f)$, we obtain
\begin{equation}
    i\Omega(f_1,f_2) = [\hat{\phi}(f_1), \hat{\phi}(f_2)] = 2i\,\text{Im}(f_1,f_2), \quad \forall f_1,f_2\in\mathcal{H}. \label{sympl-inn-prod}
\end{equation}
where the first equality is the assumed quantum realization of the classical symplectic structure. This restriction on the imaginary part of $(\cdot,\cdot)$ implies that 
\begin{equation*}
    (f_1,f_2) = \frac{1}{2}\text{Re}(f_1,f_2) + \frac{i}{2}\Omega(f_1,f_2), \quad \forall f_1,f_2\in\mathcal{H}. 
\end{equation*}
Now, by the conjugate symmetry of $(\cdot,\cdot)$, $\text{Re}(\cdot,\cdot)$ defines a positive definite metric on $\mathcal{H}$. Thus, we may specify $\text{Re}(\cdot,\cdot)$ and hence fix $(\cdot,\cdot)$ completely by requiring that $\Omega(f_1,Jf_2)$ is a positive definite metric on $\mathcal{H}$ and setting it equal to $\text{Re}(f_1,f_2)$. In other words, we can require $(\mathcal{S_{\mathbb{C}}}, J,\Omega)$ to be a K\"{a}hler space and let
\begin{equation}
    (f_1,f_2) := \frac{1}{2}\Omega(f_1,Jf_2) + \frac{i}{2}\Omega(f_1,f_2).  \label{inn-prod-asht-magn}
\end{equation}
Completion of $\mathcal{S}_{\mathbb{C}}$ in this inner product yields $\mathcal{H}$, which can be used to define the Fock space as above. To summarize, the construction of a Fock space in standard quantum field theory proceeds along the following lines.
\begin{enumerate}
    \item Take a symplectic space $(\mathcal{S}, \Omega)$ and a complex structure $J$ on it. \label{AM-construction-1}
    \item $J$ must be compatible with the symplectic structure $\Omega$ in the sense that $\Omega(f_1,Jf_2)$ defines a positive definite metric on $\mathcal{S}_{\mathbb{C}}$. Apart from this, the choice of $J$ is completely arbitrary. \label{AM-construction-2}
    \item The one-particle Hilbert space $\mathcal{H}$ on which the Fock space is based is the completion of $(\mathcal{S}_{\mathbb{C}}, (\cdot,\cdot))$, with $(\cdot,\cdot)$ as defined above \eqref{inn-prod-asht-magn}. \label{AM-construction-3}
\end{enumerate}
To illustrate this abstract construction, recall the example of a free Klein-Gordon field on Minkowski space. There, every solution of the Klein-Gordon equation can be written as a sum of positive and negative frequency solutions, i.e. $f = f^+ + f^- \quad  \forall f\in\mathcal{S}$. The complex structure $J$ can be taken \cite{Ashtekar-magnon} to be $Jf = if^+ - if^-$. This can be shown to be compatible with the symplectic structure, which is given by
\begin{equation*}
    \Omega(f_1,f_2) = \int_{\Sigma_t} \diff^3x \, (f_2\partial_tf_1 - f_1\partial_t f_2)~,
\end{equation*}
where $\Sigma_t$ is a spacelike hypersurface at time $t$. One then decomposes the field $\phi$ into creation and annihilation operators which satisfy the algebra \eqref{AM-comms}, which is realized on the Fock space $\mathcal{F}(\mathcal{H})$ over $\mathcal{H}$. 

Notice that the algorithm outlined above gives rise to a Fock space. As we will see below, in a GFT context, it is not altogether obvious if a Fock space representation exists for every GFT. Therefore, it is desirable to consider the possibility of non-Fock representations for a given theory. This question is most naturally studied in the framework of algebraic quantum field theory. Let us, therefore, review it briefly; for details, see Ref. \cite{Wald:1995yp, araki-woods}.

\subsubsection{Algebraic quantum field theory approach}
Suppose that one has constructed a Fock space $\mathcal{F(\mathcal{H)}}$ as outlined above for some choice of the triplet $(S,\Omega,J)$. On this Fock space, the classical field is realized as a self-adjoint operator-valued distribution $\hat{\phi}: \mathcal{H}\to\mathcal{F}(\mathcal{H})$, which satisfies the canonical algebra,
\begin{equation}
    [\hat{\phi}(f_1), \hat{\phi}(f_2)] = i\Omega(f_1,f_2), \quad\forall f_1,f_2\in\mathcal{H}
\end{equation}
Since $\hat{\phi}(f)$ is an unbounded operator, it is defined, in general, only on a dense domain of $\mathcal{F}(\mathcal{H})$. This means that operations such as commutation and multiplication may not be well-defined \cite[Chapter~VIII]{reed1980methods}. For this reason, it is better to work with a bounded set of operators. To this end, one can define
\begin{equation}
    W(f) := e^{i\hat{\phi}(f)}\,.
\end{equation}
These operators are bounded on $\mathcal{H}$ and, by virtue of the self-adjointness of $\hat{\phi}(f)$ and the commutation algebra above, satisfy
\begin{align}
    W(f_1)W(f_2) &= e^{-i\Omega(f_1,f_2)/2}\,W(f_1+f_2) \, , \label{weyl-relations-1}\\
    W^\dagger(f) &= W(-f) \label{weyl-relations-2}\, 
\end{align}
for all $f, f_1,f_2\in\mathcal{H}$. These equations are known as the Weyl relations. Finite linear combinations of the operators $W(f)$ form a specific example of a $*$-algebra , which is any algebra equipped with an involution operation $*$, i.e. a map on an algebra $\mathcal{U}$ satisfying
\begin{equation}
    A^{**} = A, \quad (A+B)^{*}=A^*+B^*, \quad (AB)^* = B^*A^*, \quad (\lambda A)^* = \bar{\lambda}A^*\,
\end{equation}
for all $A,B\in\mathcal{U}$ and complex numbers $\lambda$. In the case of $W(f)$ above, the star operation is the adjoint operation $\dagger$. Completion of the $*$-algebra of the $W(f)$'s in the norm on $\mathcal{F}(\mathcal{H})$ gives rise to a Banach algebra known as the Weyl algebra $\mathcal{A}$. It is a specific example of a $C^*$-algebra. 

The physical content of the quantum theory is encapsulated in the correlation functions $\braket{\hat{\phi}(f_1)\cdots\hat{\phi}(f_n)}$. Equivalently, one may consider the expectation values of Weyl algebra elements, $\braket{A}$, $A\in\mathcal{A}$, where $A = \sum_{i=1}^{n}a_iW(f_i)$ for some $a_i\in\mathbb{C}$ and $f_i\in\mathcal{H}$. Now, one can abstract away the following salient features of $\braket{A}$ that hold in any theory:
\begin{equation}
    \braket{A^*A} \geq 0, \quad \braket{I} = 1, \label{corr-conditions}
\end{equation}
where $I$ is the identity element in $\mathcal{A}$. Furthermore, $\braket{\cdot}$ is also a linear functional on $\mathcal{A}$. 

So far, we have merely reformulated the entire theory in terms of a Fock space representation of a Weyl algebra. The upshot of this reformulation is that the construction of the Weyl algebra does not depend on the choice of the complex structure $J$ that one started with \cite{Wald:1995yp}. Different choices of $J$ may lead to different Fock spaces, but the Weyl algebras they lead to are isomorphic as $C^*$-algebras. This allows us to reverse the order of things in the construction of a quantum theory as follows.

Suppose that we start with a classical theory determined by $(\mathcal{S},\Omega)$. Construct a one-particle Hilbert space $\mathcal{H}$ using a complex structure on $\mathcal{S}$ as outlined above in points \ref{AM-construction-1}-\ref{AM-construction-3} below \eqref{inn-prod-asht-magn}. On $\mathcal{H}$ define the set $\mathcal{A}(\mathcal{H})$ of complex-valued functions of finite support such that for all $f\in\mathcal{H}$, there exist elements $W(f)\in \mathcal{A}(\mathcal{H})$ that satisfy the Weyl relations \eqref{weyl-relations-1}-\eqref{weyl-relations-2} (the $\dagger$ is replaced by some abstract involution $*$ operation). Then $W(f)$ form a basis for the Weyl $*$-algebra on $\mathcal{H}$. The Weyl $*$-algebra can be turned into a $C^*$-algebra by putting a suitable norm on it; we assume this has been done. 

Define a linear map $\omega: \mathcal{A}\to\mathbb{C}$ such that for all $A\in\mathcal{A}(\mathcal{H})$,
\begin{equation}
    \omega(A^*A) \geq 0, \quad \text{and}\quad  \omega(I) = 1. \label{alg-state}
\end{equation}
It will be noticed that $\omega$, which is called an \textit{algebraic state}, satisfies the properties of quantum expectation values adumbrated above \eqref{corr-conditions}. It is more than a coincidence, for there exists the so-called Gelfand-Naimark-Segal (GNS) theorem, which asserts that given a unital $C^*$-algebra $\mathcal{A}(\mathcal{H})$ and an algebraic state $\omega$ on it, 
\begin{enumerate}
    \item there exists a representation $\pi$ of $\mathcal{A}(\mathcal{H})$ as a set of linear bounded operators on some Hilbert space $\mathcal{H}_\omega$;
    \item there exists a cyclic vector $\ket{\Psi}$ in $\mathcal{H}_{\omega}$, meaning that any state in $\mathcal{H}_{\omega}$ can be obtained by acting on the cyclic vector with a suitable linear combination of $\pi(W(f_i))$ for some $f_i\in\mathcal{H}$ and $W(f_i)\in\mathcal{A}(\mathcal{H})$;
    \item for any $A\in\mathcal{A}(\mathcal{H})$, 
    \begin{equation}
        \omega(A) = \braket{\Psi|\pi(A)|\Psi}.
    \end{equation}
\end{enumerate}
Thus, the algebraic state is the analogue of observable expectation values and as such, serves to characterize the quantum theory. While the role of a Hilbert space is now of secondary importance, if the need arises, the GNS theorem ensures the existence of a Hilbert space of quantum states for every choice of $\omega$. Different choices of $\omega$ will give rise to different, unitarily inequivalent choices for the Hilbert space. These may or may not be Fock spaces. Thus, the algebraic approach allows one to study all Hilbert space representations, Fock or non-Fock, of the quantum theory on an equal footing.  

A precise characterization of whether a representation is Fock or not can also be given. Let $\xi: \mathcal{H}\to\mathbb{C}$ be a continuous linear functional, $\mathcal{H}$ being our one-particle Hilbert space. Let $\mathcal{F}(\mathcal{H})$ be the symmetric Fock space over $\mathcal{H}$. Define an algebraic state by
\begin{equation}
    \omega_\xi(W(f)) = e^{-(f,f)/4}e^{i\text{Re}(\xi(f))} \label{fock-state}
\end{equation}
It can be shown \cite{honegger, Kegeles:2017ems, Kegeles:2018tyo} that this algebraic state gives rise to a Hilbert space representation which is equivalent to $\mathcal{F}(\mathcal{H})$ if and only if $\xi$ is bounded on $\mathcal{H}$. A heuristic explanation \cite{Kegeles:2018tyo} for this is that in a Fock space, the expectation value of the number operator $\hat{N}$ would be finite, but in the representation given by $\omega_\xi$, one has $\braket{\hat{N}} = (\xi,\xi)$, which is divergent.  

Finally, it is also useful to mention how to construct field operators $\hat{\phi}(f)$ in the algebraic formalism. For every $f\in\mathcal{H}$, if $\omega(W(tf))$ is smooth in $t\in\mathbb{R}$ (which is the case for \eqref{fock-state}), one can define a field operator $\hat{\phi}(f)$ by
\begin{equation}
    \hat{\phi}(f) = -i\partial_t\, \omega(W(tf))|_{t=0}\, , \label{field-op}
\end{equation}
where $\pi$ is the GNS representation corresponding to $\omega$; this operator is self-adjoint \cite{Kegeles:2018tyo} on its domain. With this definition, correlation functions can be obtained by
\begin{equation}
    \braket{\Psi|\hat{\phi}(f_1)\cdots\hat{\phi}(f_n)|\Psi} = (-i)^n\partial_{t_1}\cdots\partial_{t_n}\,\omega(W(t_1f_1)\cdots W(t_nf_n))|_{t_i=0\forall i}\, , \label{corr-funcs-algebraic}
\end{equation}
$\Psi$ being the cyclic vector. 

To summarize, the algebraic approach to constructing a quantum theory consists of the following steps.
\begin{enumerate}
    \item Take a symplectic space $(\mathcal{S}, \Omega)$. Construct a one-particle Hilbert space $\mathcal{H}$ as outlined above. \label{agft-1}
    \item On $\mathcal{H}$ define an abstract $C^*$-algebra $\mathcal{A}$ satisfying the Weyl relations. \label{agft-2}
    \item Define an algebraic state on $\mathcal{A}$. A useful choice is \eqref{fock-state}, which under suitable conditions will give rise to a Fock space via the GNS theorem. \label{agft-3}
    \item  Use the algebraic state to find observable expectation values. These can also be written as operator correlation functions in a Hilbert space through the GNS theorem. \label{agft-4}
\end{enumerate}

With this quick overview in mind, we want to ask: to what extent can this formalism be extended to a GFT setting? It will be noticed that it is only the choice of $\mathcal{S}$ and $\Omega$ that is tied to the theory in question. Therefore, the relevant question concerns the choice of $(\mathcal{S}, \Omega)$. It is not obvious what should be the choice of either of these objects in a GFT context. For GFTs coming from actions like \eqref{Oaction} or \eqref{action-hk-gft}, the usual choice of the space of extrema of the action does not seem to be sensible. This is because the variation of these actions leads to algebraic relations. For example, for $S_{\text{HK-GFT}}$ \eqref{action-hk-gft}, the solution space is simply $\phi=0$, which would seem to imply that there is no quantum theory corresponding to this action at all. On the other hand, we know from a path-integral perspective \eqref{pf-hk}-\eqref{eqn:npointfunctions} and its associated spinfoam interpretation that a quantum theory does undoubtedly exist. Therefore, the construction of an operator theory via steps \ref{agft-1}-\ref{agft-4} above must proceed from a choice of $\mathcal{S}$ different from the solution space of the classical theory. What should this choice be? As pointed out, a meaningful path integral can be defined for the theory. Thus, a reasonable choice of $\mathcal{S}$ is the set of fields that can enter the path integral. This is simply the space of all real-valued, right-invariant fields on $\mathrm{SU}(2)^4$. Accordingly, in the sequel, we shall take this space to be the space of physically relevant field configurations, rather than the solution space of theory. 

This settles the question of the choice of $\mathcal{S}$. For the choice of the symplectic structure $\Omega$ on $\mathcal{S}$, again the standard choice in field theory is not available in a GFT. The standard choice involves an integral over an initial-data surface \cite{ashtekar-cov, Crnkovic:1986ex}. This is evidently not possible in a GFT, for in a GFT $\mathcal{S}$ as defined above does not arise as a solution space of a set of hyperbolic partial differential equations and thus cannot be identified with initial data on a spacelike hypersurface. Thus, the choice of $\Omega$ needs to be motivated differently. Just as in the case of choosing $\mathcal{S}$, the appropriate symplectic structure is derived from the full quantum theory, as encoded in the GFT path integral \cite{Kegeles:2018tyo}. This approach is natural in the context of GFT, since the theory is fundamentally defined at the quantum level, and its classical limit—understood as the tree-level approximation of the quantum amplitudes—does not correspond to a classical gravitational theory \cite{Freidel:2005qe}.

In this sense, the Fock space construction in GFT inverts the order of steps \ref{AM-construction-1}–\ref{AM-construction-3}. More precisely, as explained in \cite{Kegeles:2018tyo}, and will be reviewed in detail below, we proceed along the following steps.
\begin{enumerate} 
\item\label{item:gft1} Starting from a background-field expansion of the GFT action, we extract from the path integral the natural structure of the GFT one-particle Hilbert space $\mathcal{H}$. 
\item\label{item:gft2} Given the configuration space $\mathcal{S}$, we define a complex structure $J$ on $\mathcal{S}$ and construct a $J$-compatible symplectic structure $\Omega$, which induces the inner product \eqref{inn-prod-asht-magn} and thereby defines $\mathcal{H}$. 
\item\label{item:gft3} With $\Omega$ identified, we proceed to the Fock quantization of the theory by following the steps \ref{agft-1}–\ref{agft-4} above. 
\end{enumerate}
The first two steps will be discussed in detail in Sec. \ref{sec:sympstructgft}, and the last step will be presented in Sec. \ref{sec:operatortheory}. 

\subsubsection{A symplectic structure for GFT}\label{sec:sympstructgft}
Let $\mu$ be a local minimum\footnote{This is a non-trivial assumption, for GFTs. Indeed, if one were to add to the action a potential term of the form $-m^2\phi^2$, this would not be the case. Although unstable \emph{mean-field} solutions are of particular interest for cosmological applications \cite{Oriti:2016qtz,Wilson-Ewing:2018mrp,Gielen:2020fgi,Marchetti:2020umh}, it is important to emphasize that within a perspective-neutral approach (that we follow here), they are only obtained after a mean-field approximation (and systematic derivative truncation) of the underlying perspective-neutral quantum equations of motion \cite{Marchetti:2020umh}. So, the classical action may still admit a local minimum
even if the resulting mean-field dynamics is unstable.} of a GFT action $S[\phi]$, i.e. $\frac{\delta S[\phi]}{\delta\phi(x)}|_{\phi=\mu}$ = 0, with $\phi:\mathrm{SU(2)}^4\to\mathbb{R}$. Then the action, expanded up to second order around $\mu$, becomes
\begin{equation}
    S[\mu+\epsilon\varphi] = S[\mu] + \frac{\epsilon^2}{2}\int \diff g_I \diff h_I\,\varphi(h_I) \left[\frac{\delta^2 S[\phi]}{\delta \phi(g_I)\phi(h_I)}\right]_{\phi=\mu}\varphi(g_I)\,,
\end{equation}
and thus the generating functional $Z[J] = \int D\phi\exp{(-S[\phi] + \int J\phi)}$ is rendered into
\begin{equation}
    Z[J] = e^{-S[\mu]+\int dg_I\, J(g_I)\mu(g_I)} \int D\varphi\, \exp{\left[-\frac{1}{2}\int \diff g_I\, \varphi(g_I) K_{\mu}(\varphi)(g_I) + \int \diff g_I\, J(g_I)\varphi(g_I)\right]}\, , \label{gen-fun-min}
\end{equation}
where we have absorbed $\epsilon$ into a redefinition of $\varphi$ and defined an operator $K_{\mu}$ on the space of all group fields using the second functional derivative of the action:
\begin{equation}
    K_{\mu}(\varphi)(g_I) = \int \diff h_I\left[\frac{\delta}{\delta\phi(g_I)}\frac{\delta}{\delta\phi(h_I)}S[\phi]\right]_{\phi=\mu}\varphi(h_I) \label{operator-K}\, . 
\end{equation}
We can in fact extend it to any function on $SU(2)^4$, even complex-valued functions. Since $\mu$ is a local minimum of the action, $K_{\mu} > 0$ and therefore, the functional integral above is a standard Gaussian integral, which can be evaluated, giving 
\begin{equation}
    Z[J] = e^{\int \diff g_I\, J(g_I)\mu(g_I)} e^{\frac{1}{2}\int \diff g_I J(g_I)K^{-1}_{\mu}(J)(g_I)}\, , \label{gen-fun-min-eval}
\end{equation}
where the operator $K^{-1}_{\mu}$ is the Green function (inverse) of $K_{\mu}$. Now if this generating functional is used to evaluate correlation functions, one would obtain, for instance, 
\begin{equation}
    \braket{\phi(g_I)\phi(h_I)} = \mu(g_I)\mu(h_I) + \frac{1}{2}K^{-1}_{\mu}(g_I,h_I)\, \label{corr-mu},
\end{equation}
$K^{-1}_{\mu}(g_I,h_I)$ being the kernel of $K^{-1}_{\mu}$. In other words, \eqref{gen-fun-min-eval} defines a ``non-interacting theory'' of a field $\phi$ with a classical value $\mu$ and Green function $K^{-1}_{\mu}$. The preceding equation suggests that if in a yet-to-be-found operator theory, $\phi$ can be decomposed into a set of creation and annihilation operators $\hat{A}^\dagger(g_I)$ and $\hat{A}(g_I)$, then we must have
\begin{equation}
    [\hat{A}(g_I), \hat{A}^\dagger(h_I)] = K^{-1}_{\mu} (g_I,h_I). \label{comm-K}
\end{equation}
Recalling the steps leading up to \eqref{inn-prod-asht-magn}, the preceding equation suggests that the inner product on our one-particle Hilbert space, whatever it may be, must come from $K^{-1}_{\mu}$. In fact, note that the operator $K^{-1}_{\mu}$ defines a natural inner product on the space $\mathcal{C}$ of \textit{complex-valued} right-invariant functions on $\mathrm{SU}(2)^4$, namely
\begin{equation}
    (f_1, f_2)_{K^{-1}_{\mu}} \equiv (f_1, K^{-1}_{\mu}f_2) = \int \diff h_I\ \Bar{f_1}(h_I)K^{-1}_{\mu}(f_2)(h_I),\qquad \forall\, f_1, f_2 \in \mathcal{C}\, \label{inner-prod-C} ,
\end{equation}
where the integration on the right is performed using the standard Haar measure. This is anti-linear and linear in the first and second arguments by construction, and positivity follows from the positivity of $K^{-1}_{\mu}$ \cite{Kegeles:2018tyo}. Thus, \eqref{inner-prod-C} defines an inner product on $\mathcal{C}$. Taking $\mathcal{C}$ to be the space of smearing functions for $\hat{A}$ and $\hat{A}^\dagger$ regarded as operator-valued distributions, the smeared version of \eqref{comm-K} would be
\begin{equation}
    [\hat{A}(f_1), \hat{A}^\dagger(f_2)] = (f_1,f_2)_{K^{-1}_\mu}\quad \forall f_1,f_2\in\mathcal{C}\,.
\end{equation}
In other words, one can regard the completion of $\mathcal{C}$ in $(\cdot,\cdot)_{K^{-1}_\mu}$ as the one-particle Hilbert space $\mathcal{
H}$ of our theory, and thus complete step \ref{item:gft1} above. Since we know what $\mathcal{S}$ is—namely, the space of real-valued, right-invariant functions on $\mathrm{SU}(2)^4$—and have a reasonable guess for $\mathcal{H}$, we can now address step~\ref{item:gft2} of the construction: identifying the complex and symplectic structures, $J$ and $\Omega$ respectively, that give rise to $\mathcal{H}$.

Recall that the inner product on the one-particle Hilbert space has to be compatible with the symplectic structure in the sense of \eqref{sympl-inn-prod}. This invites us to define \cite{Kegeles:2018tyo} a symplectic product on $\mathcal{C}$:
\begin{equation}
    \Omega(f_1,f_2) := 2\text{Im}(f_1, f_2)_{K^{-1}_{\mu}} \label{sympl-gft}
\end{equation}
This evidently satisfies all the properties of a symplectic product (i.e. it is real bilinear, antisymmetric and nondegenerate). One now contends \cite{Kegeles:2018tyo} that this is the symplectic structure of the classical GFT defined by the GFT action $S[\phi]$. On $\mathcal{S}$, which, recall, is the space of \textit{real-valued} right-invariant functions on $\mathrm{SU}(2)^4$, the symplectic product is trivial, $\Omega(f_1,f_2)=0$ for all $f_1,f_2\in\mathcal{S}$. However, on $\mathcal{C}$, $\Omega$ acts nontrivially. We can put a complex structure $J$ on $\mathcal{S}$ that turns it into $\mathcal{C}$ and is compatible with $\Omega$. Indeed, take $Jf = if$ for every $f\in \mathcal{S}$. Thus for every $f\in\mathcal{C}$, $Jf = J(\text{Re}(f) + i\text{Im}(f)) = i\text{Re}(f) - \text{Im}(f) = if$. Therefore, for every $f_1,f_2\in\mathcal{C}$, we have
\begin{align*}
    \Omega(f_1,Jf_2) = 2\text{Im}(f_1,Jf_2)_{K^{-1}_{\mu}} &= 2\text{Im}(f_1,if_2)_{K^{-1}_{\mu}} \\
    &= -i[(f_1,if_2)_{K^{-1}_{\mu}} - \overline{(f_1,if_2)}_{K^{-1}_{\mu}}] \\
    &= [(f_1,f_2)_{K^{-1}_{\mu}} + \overline{(f_1,f_2)}_{K^{-1}_{\mu}}] = 2\text{Re}(f_1,f_2)_{K^{-1}_{\mu}},
\end{align*}
which is positive definite on $\mathcal{C}$, and it follows that (cf. \eqref{inn-prod-asht-magn}).
\begin{equation}
    (f_1,f_2)_{K^{-1}_{\mu}} = \frac{1}{2}\Omega(f_1,Jf_2) + \frac{i}{2}\Omega(f_1,f_2)\,.
\end{equation}
This completes step~\ref{item:gft2} of the construction. It also justifies our initial choice of $\mathcal{H}$, as it emerges from a specific choice of $J$ and $\Omega$ on $\mathcal{S}$—the space of physically relevant field configurations. We now turn to step~\ref{item:gft3}.

\subsubsection{Operator theory}\label{sec:operatortheory}
Now that we are in the possession of a symplectic structure and a one-particle Hilbert space, we can accomplish the first two steps of the algebraic approach (cf. steps \ref{agft-1}-\ref{agft-4} below \eqref{corr-funcs-algebraic}). We start by picking the algebraic state \cite{Kegeles:2018tyo}
\begin{equation}
    \omega_{\mu}(W(f)) = e^{-\frac{1}{4}(f,f)_{K^{-1}_\mu}}e^{i\text{Re}(\mu(f))}\, , \label{algebraic-state-gft}
\end{equation}
where 
\begin{equation}
    \mu(f) = \int\diff g_I\, f(g_I)\mu(g_I)\,.
\end{equation}
With this choice, and the use of \eqref{corr-funcs-algebraic}, the correlation functions coincide with the path integral result \eqref{corr-mu}. As an illustration, consider the two-point function. We have
\begin{align}
    \braket{\Psi|\hat{\phi}(f_1)\hat{\phi}(f_2)|\Psi} &= -\partial_{t_1}\partial_{t_2} \,\omega_\mu(W(t_1f_1)W(t_2f_2))|_{t_1,t_2=0} \nonumber\\
    &= -\partial_{t_1}\partial_{t_2}\, \omega_{\mu}(W(t_1f_1+t_2f_2))e^{-i\Omega(t_1f_1,t_2f_2)/2}|_{t_1,t_2=0}\nonumber\\
    &= -\partial_{t_1}\partial_{t_2}\exp\left[ -\frac{t^2_1}{4}(f_1,f_1)_{K^{-1}_\mu} - \frac{t_2^2}{4}(f_2,f_2)_{K^{-1}_\mu} - \frac{t_1t_2}{2}(f_1,f_2)_{K^{-1}_\mu}\right] \nonumber \\
    &\quad \times \exp\left[it_1\text{Re}(\mu(f_1))+ it_2\text{Re}(\mu(f_2))\right]\exp[-it_1t_2\Omega(f_1,f_2)/2]|_{t_1=0,t_2=0}\nonumber \\
    &= \frac{1}{2}(f_1,f_2)_{K^{-1}_\mu} + \text{Re}(\mu(f_1))\text{Re}(\mu(f_2)). 
\end{align}
For real $f_1,f_2$, this is simply the smeared version of \eqref{corr-mu}. More generally, one can show \cite{Kegeles:2018tyo} that 
\begin{align}
    \partial_{t_1}\cdots\partial_{t_n}\,\omega(&W(t_1f_1)\cdots W(t_nf_n))|_{t=0} = \nonumber \\
    &\int \diff g_{I_1}\cdots dg_{I_n}\,f(g_{I_1})\cdots f(g_{I_n})\frac{\delta}{\delta J(g_{I_1})}\cdots \frac{\delta}{\delta J(g_{I_n})}Z[J]|_{J=0}
\end{align}
where $Z[J]$ is the generating functional \eqref{gen-fun-min-eval} for the theory around $\mu$. This shows that we have succeeded in constructing a successful operator theory for a general GFT action $S[\phi]$. 

Two points should be noted about this construction. First, it depends explicitly on the local minimum $\mu$ around which one expands the action. Thus, the algebraic approach to GFT enables one to construct an operator theory in the vicinity of every local minimum of the action. The different theories arising in this way can be thought of as the \qmarks{phases} of a GFT \cite{Kegeles:2018tyo}.

Secondly, as we saw above, whether an operator theory has a Fock space representation or not depends on the functional $\mu(f)$ entering the definition of the algebraic state \eqref{algebraic-state-gft}. Only when $\mu$ is bounded on $\mathcal{H}$ can the correlation functions be understood as Fock-space expectation values, the cyclic vector $\ket{\Psi}$ being the Fock vacuum. 

\subsubsection{Fock space for HK-GFT}
We now have all the tools in hand to fulfill our original quest of finding a suitable Hilbert space representation for the GFT of the HK model. The action is 
\begin{equation}
    S_{\text{HK-GFT}}[\phi] = \frac{1}{2}\int \diff g_I \,\phi^2(g_I)\, ,
\end{equation}
subject to the condition that the field $\phi$ is invariant under permutations of all its arguments. This implies that the space $\mathcal{C}$ introduced above will be subject to this restriction too. That is, $\mathcal{C}$ will now be the space of right-invariant complex-valued functions on $\mathrm{SU}(2)^4$ that are also invariant under permutations of their arguments. 

Varying the above action once reveals that $\phi = 0$ is the unique minimum of the action. Thus, there will be one, and only one, operator theory arising out of the algebraic machinery. Moreover, since the functional $\mu$ is now zero as well, the GNS representation will be equivalent to a Fock space over the one-particle Hilbert space $\mathcal{H}$. To see the structure of $\mathcal{H}$, vary the action twice to find the operator $K$ \eqref{operator-K}. One finds that it is the identity. Therefore, the operator $K^{-1}_\mu$ also acts as the identity on $\mathcal{C}$ so the inner product \eqref{inner-prod-C} reduces to the $L^2$ inner product on $\mathrm{SU}(2)^4$ with respect to Haar measure,
\begin{equation}
    (f_1,f_2)_{K^{-1}_{\mu}} \equiv (f_1,f_2) = \int \diff g_I\, \bar{f}_1(g_I)f_2(g_I), \quad \forall f_1,f_2\in\mathcal{C\,.}
\end{equation}
Therefore, the one-particle Hilbert space $\mathcal{H}$ is simply the space of right-invariant, permutation-invariant and complex-valued square-integrable functions on $\mathrm{SU}(2)^4$. With this, the correlation functions \eqref{corr-funcs-algebraic} turn out to be
\begin{equation}
    \braket{0|\hat{\phi}(f_1)\cdots\hat{\phi}(f_n)|0} = \frac{1}{2}\sum_{\text{all pairings}}\prod_{i\neq j}^{n}(f_i,f_j)\, \label{corr-alg},
\end{equation}
which, up to a factor of 2 (which can be absorbed into a redefinition of the symplectic product), agrees with the path integral result \eqref{eqn:npointfunctions} for the HK model. Here we have written $\ket0$ for the cyclic vector of the GNS representation, in anticipation of the fact that the latter is equivalent to the Fock space $\mathcal{F}(\mathcal{H})$. 

By virtue of the Weyl relations \eqref{weyl-relations-1}, the field operators $\hat{\phi}(f)$ satisfy
\begin{equation}
    [\hat{\phi}(f_1),\hat{\phi}(f_2)] = i\Omega(f_1,f_2) = 2i\text{Im}(f_1,f_2), \quad \forall f_1,f_2\in\mathcal{H}.
\end{equation}
Due to the self-adjointness of $\hat{\phi}(f)$, it is only \textit{real}-linear, i.e. $\hat{\phi}(\lambda f) =\lambda\hat{\phi}(f)$ for $\lambda \in \mathbb{R}$ but not $\lambda$ imaginary\footnote{Assume  $\hat{\phi}(if)=i\hat{\phi}(f)$. Then by the definition of the adjoint, $\hat{\phi}(if)^{\dagger}= -i\hat{\phi}(f)^{\dagger}$. But self-adjointness implies that $\hat{\phi}(f)^{\dagger}=\hat{\phi}(f)$ and $\hat{\phi}(if)^{\dagger}=\hat{\phi}(if)$, and so $i\hat{\phi}(f)=-i\hat{\phi}(f)$, which is a contradiction.}. Thus, $\hat{\phi}(f)$ and $\hat{\phi}(if)$ are independent objects. This allows us to construct creation and annihilation operators on $\mathcal{F}(\mathcal{H})$,
\begin{align}
    \hat{A}(f) &= \frac{1}{2}(\hat{\phi}(f) + i\hat{\phi}(if)) \\
    \hat{A}^\dagger(f) &= \frac{1}{2}(\hat{\phi}(f) - i\hat{\phi}(if))\, .
\end{align}
Real linearity of $\hat{\phi}(f)$ implies that $\hat{A}(f)$ is anti-linear and $\hat{A}^\dagger(f)$ is linear. Furthermore, the commutation algebra for $\hat\phi(f)$ implies that
\begin{equation}
    [\hat{A}(f_1),\hat{A}(f_2)]=[\hat{A}^\dagger(f_1),\hat{A}^\dagger(f_2)] = 0, \quad [\hat{A}(f_1),\hat{A}^\dagger(f_2)] =  (f_1,f_2)\, , \label{canon-comm-gft}
\end{equation}
as expected. Thus $\hat{A}(f)$ annihilates the cyclic vector $\ket{0}$, as can be directly verified:
\begin{align*}
    \braket{0|\hat{A}^{\dagger}(f)\hat{A}(f)|0} &= \frac{1}{4}\left(\braket{0|\hat{\phi}(f)^2|0} + \braket{0|\hat{\phi}(if)^2|0} + i\braket{0|[\hat{\phi}(f),\hat{\phi}(if)]|0}\right)\\
    &= \frac{1}{2}\left(-\partial_t^2\omega(W(tf))|_{t=0} - \partial_t^2\omega(W(itf))|_{t=0}  -  2(f,f)\right) \\
    &= \frac{1}{2}\left((f,f) + (f,f) - 2(f,f)\right) = 0 \Rightarrow \hat{A}(f)\ket{0} = 0\, .
\end{align*}

Thus, we have succeeded in constructing a Fock space representation of the GFT of the HK model. Interestingly, the algebra \eqref{canon-comm-gft} of the creation and annihilation operators above is identical to the algebra \eqref{canon-comm-complex} imposed upon a complex-valued field and its conjugate in Section \ref{sec:gft-fock}. This shows that the GFT Fock space $\mathcal{F}_{\text{GFT}}$ postulated in Section \ref{sec:gft-fock} is nothing but the Fock space $\mathcal{F}(\mathcal{H})$ that has been systematically constructed above. It might seem paradoxical at first that a Fock space for a theory defined by an action on real-valued fields, i.e. $S_{\text{HK-GFT}}$, should be the same as the Fock space which is described by a complex-valued field and its conjugate, as in Section~\ref{sec:gft-fock}. But first of all, the interpretations of the objects composing the canonical commutation algebra in the complex case \eqref{canon-comm-complex} and in the present case \eqref{canon-comm-gft} are different. In the former, $\phi, \phi^\dagger$ are field operators, introduced to push an analogy with condensed matter physics, while here these are creation and annihilation operators. Furthermore, and more importantly, we must bear in mind that $\mathcal{F}_{\text{GFT}}$ was simply postulated. In particular, the canonical commutation algebra \eqref{canon-comm-complex} between a complex field and its conjugate did not arise out of a quantization of a symplectic structure. But \eqref{canon-comm-gft} does arise in this way, and is seen to be the same as \eqref{canon-comm-complex} post facto. Therefore, one may regard the symplectic structure \eqref{sympl-gft}, $K^{-1}_\mu=\text{id}$ considered above as the classical structure underlying the abstract algebra \eqref{canon-comm-complex}. 

One caveat should be kept in mind, however. Strictly speaking, $S_{\text{HK-GFT}}$ was written down with the assumption that the fields entering it are permutation-invariant in their arguments. This symmetry, as we have seen, is also reflected in the Fock space for the GFT of the HK model. In particular, all the function spaces used above have this property. Thus, as far the Fock space for the GFT of the HK model is concerned, the above arguments show that it is equivalent to the permutation-invariant part of the Fock space $\mathcal{F}_{\text{GFT}}$ considered in Section \ref{sec:gft-fock}. Nevertheless, it is worth observing that if one drops the assumption of permutation invariance in the function spaces $\mathcal{C}$ and $\mathcal{H}$ above, one will end up with a Fock space for a non-interacting action $\int \phi^2$ without the assumption of permutation-invariance on $\phi$; this Fock space will be equivalent to the GFT Fock space postulated in Section~\ref{sec:gft-fock}. 

\begin{figure}
    \centering
    \begin{tikzpicture}[
    node distance=2cm,
    every node/.style={rectangle, draw, minimum width=2.5cm, minimum height=1cm, text centered, font=\sffamily},
    arrow/.style={-latex, thick},
    circled/.style={draw, circle, minimum width=2.5cm, minimum height=2.5cm}
    ]

\node[circled] (hk) {Classical HK model};
\node (gft) [below=1cm of hk] {non-interacting GFT};
\node (path) [below left=1cm of gft] {Path integral representation};
\node (fock) [below right=1cm of gft] {Fock representation};
\node (spinfoams) [below=1cm of path] {Spinfoams};
\node (lqg) [below=1cm of fock] {Canonical quantization};

\draw[arrow] (hk) -- node[right,text width=4.0cm,align=center,draw=none] {GFT quantization} (gft); 
\draw[arrow] (gft) -- (path);
\draw[arrow] (gft) -- (fock);
\draw[arrow] (path) -- (spinfoams);
\draw[arrow] (fock) -- (lqg);

\end{tikzpicture}
    \caption{Various connections explored in this paper.}
    \label{fig:enter-label}
\end{figure}
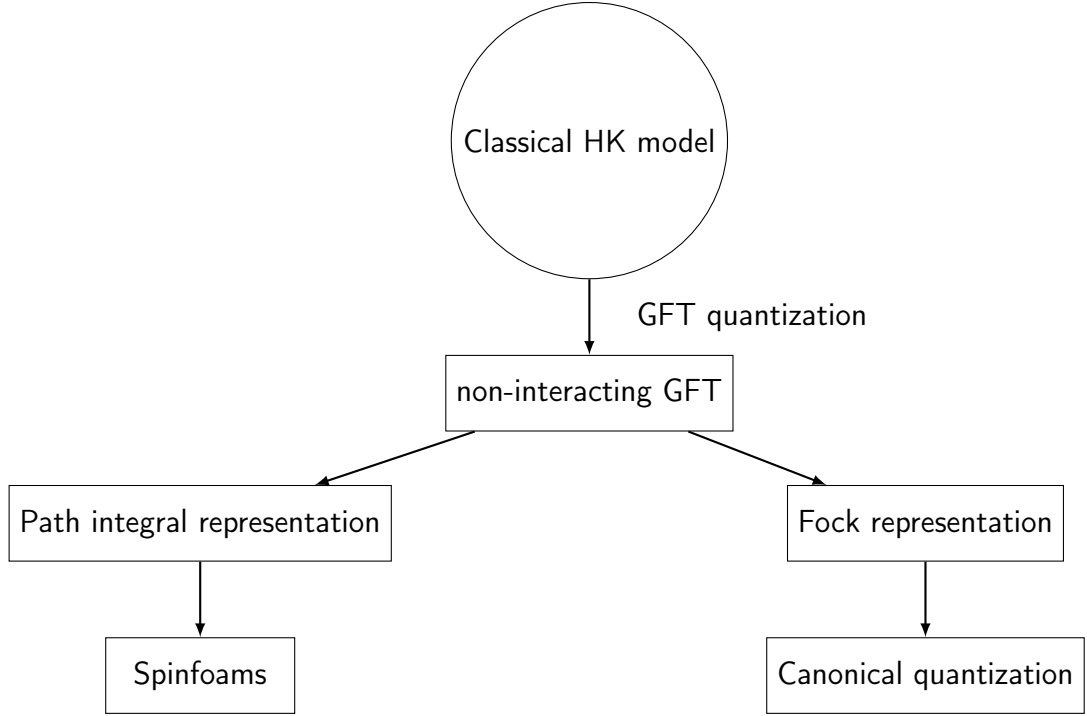

In summary, we have constructed a Fock space for a non-interacting GFT (and hence for the GFT of the HK model);  we have found that this Fock space, within the prescription presented above, is unique (since there is one, and only one, minimum for a non-interacting GFT);  we have also seen that this Fock space is equivalent to the one postulated in Section \ref{sec:gft-fock}. Therefore, the (permutation-invariant part of) the latter is the the Fock space for the GFT of the HK model, and the correlation functions \eqref{corr-alg} provide a rigorous definition within the context of GFT of the formal canonical product \eqref{informal-inn-prd}.  These conclusions, together with the fact that the HK GFT model of equation \eqref{action-hk-gft} provides a completion of the HK spinfoam model in \cite{Bojowald:2009im}, complete the link between the canonical and the spinfoam quantization of the HK model using the GFT formalism (see Fig \ref{fig:enter-label}).

\section{Conclusions and discussion}\label{sec:conc}

We presented a Group Field Theory (GFT) quantization of the HK model. This was achieved by using the existence of a preferred direction within the HK model to appropriately constrain the Ooguri model \cite{Ooguri:1992eb}. The resulting constrained action \eqref{action-hk-gft}, after the imposition of the constraint, yields an effectively non-interacting GFT. Such a non-interacting GFT describes the physics of non-dynamical quantum $3-$geometries, thereby providing a quantization of the HK model. 

From the path-integral perspective, we demonstrated that the HK GFT is a completion of the HK spinfoam model constructed in \cite{Bojowald:2009im}. Furthermore, we showed through an algebraic construction that there is a  unique Fock space for the model. Interestingly, this Fock space coincides with the one proposed in \cite{Oriti:2013aqa} and considered in Section \ref{sec:gft-fock}, which can thus be identified as a kinematical GFT Fock space. 

Within this Fock space, we demonstrated that amplitudes between physical boundary states, as described from a canonical  quantization perspective, are non-zero (and trivial) only for identical boundary states. This result coincides with that obtained within a canonical quantization of the HK model, thereby confirming that the GFT quantization of the HK model provides physically equivalent answers to those expected from a canonical LQG quantization. Since the GFT HK model also provides a completion of the HK spinfoam model of \cite{Bojowald:2009im}, it offers a clear connection between the canonical and spinfoam quantizations, thus serving as an example of how GFTs can bridge formally distinct but potentially equivalent quantization schemes.

\myparagraph{Canonical constraints in GFTs.}
As reviewed in Section \ref{sec:intro-gft}, GFTs offer a radically different quantization of gravitational theories. For this reason, it is often difficult to identify which objects at the level of the microscopic GFT correspond (in an appropriate limit) to quantities in the continuum macroscopic theory. A crucial example of this is the diffeomorphism symmetry or, equivalently, from a canonical perspective, the constraints associated with it. Since GFTs lack a continuum spacetime structure, the question of how diffeomorphisms are represented in GFTs remains open (see however \cite{Baratin:2011tg} for an attempt to address the issue in $3$d Riemannian gravity and $\mathrm{BF}$ theories). This work provides a first step towards answering this question by identifying a non-interacting GFT with the quantum version of the HK model, a background-independent theory lacking, from a canonical perspective, a Hamiltonian constraint. 

This suggests that the Hamiltonian constraint is encoded in GFT interactions (as alluded to in \cite{Oriti:2013aqa}), and possibly also in a non-trivial quadratic term 
describing  the ``propagation'' of geometric and matter data between neighboring $4$-simplices. It paves the way for a systematic analysis of the problem by perturbatively moving away from the HK model in theory space. This could be achieved by allowing the quantity $B^i_{\alpha\beta}\Tilde{u}^\alpha$ to be non-zero, but of order $\epsilon\ll 1$. This could be implemented at the level of the GFT action \eqref{action-hk-gft} by allowing a weak imposition of the constraint the spirit of \cite{Asante:2020qpa,Asante:2021zzh}. At the canonical level, this is expected to change the constraint structure, and one could compare such change to that of the effective GFT action, which is expected to move perturbatively away from a purely non-interacting theory.

\myparagraph{Topology change and physical inner product in LQG.} 
One question that we can ask of the model is whether it allows topology changing amplitudes. We argued in Section \ref{sec:4.2}, HK-GFT does not allow geometry change \footnote{Not only should the spin networks in an inner product have diffeomorphic graphs, but the spin data on them must be identical too; spin data encode the geometry of space represented by a spin network.}. Thus, a fortiori, it cannot alone topology change --- as one would expect from the properties of the classical theory.   

This  is  not surprising if one remembers the fact that we are working with a non-interacting GFT. For a field theory with interactions, one can expand the partition function in powers of the coupling constant in the interaction term. As is well-known from standard quantum field theory, the power of the coupling constant at each order counts the number of loops present in the Feynman diagrams at that order. As argued in \cite{Freidel:2005qe}, in a GFT, these loops can be interpreted as the addition of a handle to a manifold, and thus, addition of loops describes topology change. Since here we have a non-interacting GFT, there are no loops, hence no handles or topology change. 

Related to the question of topology change is a proposal presented in \cite{Freidel:2005qe} for constructing the physical scalar product in loop quantum gravity. It is argued there that given two spin network functions (which live in the kinematical Hilbert space of loop quantum gravity), their physical inner product should be defined as a sum over all tree-level GFT Feynman graphs interpolating the two spin networks. Tree-level diagrams do not involve loops, which as hinted above are related to topology change. Now, in our model, one does not have geometry change, let alone topology change. Thus this proposal for the physical inner product in LQG should apply to the HK model. This is trivially true, for in the absence of interactions, there are no tree-level graphs, and thus a ``sum over all tree-level graphs'' should reduce simply to a delta function which is nonzero only when the two spin network graphs bounding the tree-level graphs are identical. This is precisely what the physical inner product in a canonical LQG quantization of the HK model achieves, as we saw in Section \ref{sec:hk}. Let us emphasize that our results do not clarify  whether it is the quantum GFT amplitude or the classical one that gives the appropriate physical inner product, but they allow a study the problem systematically, by perturbatively moving away from the HK model in theory space.

\myparagraph{Adding matter fields.}
Finally, one may consider extensions of the model presented here  by including matter degrees of freedom. Besides being relevant for representing physically realistic systems, the introduction of matter data can facilitate a relational \cite{Rovelli:1990ph,Dittrich:2005kc,Tambornino:2011vg,Goeller:2022rsx} description of the GFT system. This is an especially pressing issue in GFTs, since the lack of spacetime structures at the microscopic level requires any dynamics and localization to be defined in relational terms \cite{Marchetti:2020umh,MarchettiWilsonEwing2024}. Indeed, most of the emergent continuum physics extracted from GFTs is described within a physical frame composed of $4$ minimally coupled, massless and free (MCMF) scalar fields \cite{Gielen:2018xph,Marchetti:2021gcv,Jercher:2023nxa,Jercher:2023kfr,Gielen:2023szb, Oriti:2023yjj}, or a single MCMF scalar field clock (see \cite{Gielen:2021vdd} for a curvature clock instead), in homogeneous and isotropic settings \cite{Oriti:2016qtz,Wilson-Ewing:2018mrp,Marchetti:2020umh,Gielen:2020fgi}.  

The way that matter fields can be coupled to a GFT action follows the general procedure outlined in Section \ref{sec:intro-gft}, and it is therefore based on the construction of a GFT action that produces amplitudes that can be matched to a simplicial matter-gravity path-integral (see \cite{Li:2017uao} for an explicit scalar field example). In order to do this, we make the assumption that the symmetries of a classical matter action are also present when matter is coupled to a GFT \cite{Oriti:2016qtz}.  
For instance, a massless scalar field minimally coupled to the Einstein-Hilbert action has shift and sign-reversal symmetries. Requiring a coupled GFT-scalar-field action to also show these symmetries is thus a minimal requirement for a scalar field to be coupled to a GFT action  
\cite{Gielen:2016dss, Oriti:2016qtz}. 

 As we saw in Section \ref{sec:hk}, there are at least three distinct ways of coupling massless scalar fields to the HK model. The first one involves two fields only one of which is shift-symmetric. So for this model, one could add two scalar fields to the GFT action \eqref{action-hk-gft} in the same way. In other words, one would now have an action of the form
\begin{equation}
    \int \diff g_u \diff g_v \diff \chi_u \diff \chi_v \diff \pi_u \pi_v\, \phi(g_u,\chi_u,\pi_u)\,K(g_u,g_v,\chi_u-\chi_v, \pi_u, \pi_v)\,\phi(g_v,\chi_v,\pi_v)\, .
\end{equation}
Here $g_u = (g_{u_1}, g_{u_2}, g_{u_3}, g_{u_4})$ and similarly for $g_v$; $u$ and $v$ denote distinct vertices/tetrahedra which the GFT fields $\phi(g_u)$ and $\phi(g_v)$ respectively label; the subscripts on the scalar fields $\chi$ and $\pi$ stand for the vertices/tetrahedra they are smeared over. The kinetic kernel $K$ depends on the difference of the $\chi$ fields, since they are shift-symmetric in the classical HK model \eqref{eq2.7}.  
As for the other two prescriptions for adding a scalar field to the HK model, a non-interacting GFT does not suffice to describe those models in the first place, since they change the non-dynamical canonical structure of the HK model on which we based our arguments to describe the model by means of a non-interacting GFT. 

\begin{acknowledgements}
This work was supported by the Natural Science and Engineering Research Council of Canada. L.M.~acknowledges support from the Atlantic Association for Research in Mathematical Sciences and from the Okinawa Institute of Science and Technology Graduate University. This project/publication was also made possible through the support of the ID\# 62312 grant from the John Templeton Foundation, as part of the \href{https://www.templeton.org/grant/the-quantum-information-structure-of-spacetime-qiss-second-phase}{\textit{`The Quantum Information Structure of Spacetime'} Project (QISS)}. We thank Steffen Gielen, Daniele Oriti, and Edward Wilson-Ewing for helpful discussions. We also thank an anonymous referee for valuable comments which helped significantly improve the paper.
\end{acknowledgements}  

\appendix

\section{$\mathrm{BF}$ theory, spinfoams and all that}\label{sec:appC}
In this Appendix, we will review $\mathrm{BF}$ theory, its discretization on triangulated manifolds and sketch the derivation of the subsequent spinfoam amplitudes. 

We start with a principal $\mathrm{SU(2)}$ bundle $P$ over a manifold $M$ and define on it a connection one-form $A$ and a two-form $B$. Then on $M$, the action for a $\mathrm{BF}$ theory is
\begin{equation}
    S_{\mathrm{BF}} = \int_{M} \diff^4x\, \text{tr}(B\wedge F)\, ,
\end{equation}
where $F = A\wedge dA + A\wedge A$ is the curvature of the connection $A$. This entails that the path integral for this theory must be
\begin{align}
    Z_{\mathrm{BF}}(M) = \int DA DB \,\exp{\left(i \int_M \diff^4x\,\text{tr}(B\wedge F) \right)} = \int DA\, \delta(F) \label{eqc1}\, ,
\end{align}
where we have used the identity $\delta(x) = \int dp\,e^{ipx}$ to integrate out $B$. 

Our task is to make sense of this path integral by picking a triangulation $C_M$ of $M$, and discretizing the $\mathrm{BF}$ theory on it. This is akin to making sense of a quantum mechanical path integral by discretizing its path. To this end, we resort to the so-called dual 2-skeleton $C^*_M$ of $C_M$. It is defined by replacing every 4-simplex in $C_M$ with a vertex, every 3-simplex (tetrahedron) with an edge, and every 2-simplex (triangle) with a 2-simplex (which may have any number of edges). Then by discretization of the $\mathrm{BF}$ theory we mean that we define the connection $A$ and the two-form $B$ on elements in this dual 2-skeleton. As in the case of a GFT, this can be done in three equivalent pictures: the group representation, the spin representation and the Lie algebra representation. In the group representation, for instance, the $B$ field is smeared over the faces in $C^*_M$, while the connection $A$ is replaced by holonomies around edges in $C^*_M$; the curvature concentrated at a face is defined as the holonomy of the edges surrounding it. The action on $C_M$ thus becomes
\begin{equation}
    S_{\mathrm{BF}}(C_M) = \sum_{f^*\in C^*_M}\text{tr}(B_{f^*} H_{f^*})\, ,
\end{equation}
where $B_{f^*}$ is the $B$ field smeared over the face $f^*$ and $H_{f^*}$ is the holonomy along the edges surrounding $f^*$. The discretized path integral is then
\begin{equation}
    Z_{\mathrm{BF}}(C_M) = \int \prod_{e^*\in C^*_M} \diff g_{e^*} \prod_{f^*\in C^*_M} e^{i\text{tr}(B_{f^*}H_{f^*})}\,,
\end{equation}
where $g_{e^*}$ is the holonomy along the edge $e^*$. If one switches to the Lie algebra representation, then the $B_{f^*}$ are interpreted as normal vectors to the faces $f$ to which the dual faces $f^*$ are dual \cite{Engle:2007qf, Rovelli:2014ssa}.  
Integrating out $B$, one obtains
\begin{equation}
    Z_{\mathrm{BF}}(C_M) = \int \prod_{e^*\in C^*_M} \diff g_{e^*}\prod_{f^*\in C^*_M}\delta(H_{f^*},1)\, .
\end{equation}
Denoting by $e^*_1(f^*), \ldots, e^*_n(f^*)$ the edges surrounding $f^*$,  
\begin{equation}
    H_{f^*} = g_{e^*_1(f^*)}\cdots g_{e^*_n(f^*)}\, ,
\end{equation}
where an orientation on the edges complying with the direction in which one traverses around the face is understood. Therefore,
\begin{equation}
    Z_{\mathrm{BF}}(C_M) = \int \prod_{e^*\in C^*_M} \diff g_{e^*} \prod_{f^*\in C^*_M}\delta(g_{e^*_1(f^*)}\cdots g_{e^*_n(f^*)}, 1)\, . 
\end{equation}
But this is equivalent to  \eqref{eq3.20}, if one remembers the fact that a product over dual edges surrounding a dual face in $C^*_M$ corresponds to a face shared by tetrahedra to which the edges are dual. Furthermore, using the harmonic theory of $\mathrm{SU(2)}$, we can write the preceding amplitude in the spin representation as well. This would of course coincide with the spin representation of the GFT amplitude in Section \ref{sec:gft-amps}. Therefore, a GFT partition function does indeed furnish a generating functional for spinfoam amplitudes.

Finally, it is worth pointing out that there is only a short step one needs to take in going from $\mathrm{BF}$ theory to gravity, which explains the utility of the former. There are a number of distinct ways of doing this. For example, if one imposes the constraint $B = e\wedge e$ in a $\mathrm{BF}$ theory with gauge group $SO(3,1)$, then one obtains the Palatini formulation of general relativity. A slightly different constraint leads to the Plebanski-Holst formulation of general relativity \cite{Engle:2007qf, Rovelli:2014ssa}. The derivation of a spinfoam amplitude for such constrained theories typically involves imposing the relevant constraints on the vertex amplitude of the associated $\mathrm{BF}$ theory. The resulting amplitude depends on the manner in which the constraints are imposed. Two famous examples include the Barret-Crane model and the EPRL model \cite{Engle:2007qf}. There have also been attempts to obtain more general spinfoam amplitudes of which the Barret-Crane and EPRL models are special cases (see \cite{Alexandrov:2011ab} and the references therein).

\bibliography{main.v3}
    
\end{document}